\def\be{\begin{equation}}
\def\te{\end{equation}}
\def\bea{\begin{eqnarray}}
\def\nn{\nonumber}
\def\tea{\end{eqnarray}}

\def\a{\alpha}
\def\b{\beta}

\def\d{\delta}
\def\e{\epsilon}
\def\g{\raisebox{.4ex}{$\gamma$}}

\def\m{\mu}
\def\n{\nu}
\def\o{\omega}

\def\t{\tau}

\def\D{\Delta}

\def\O{\Omega}

\def\S{\Sigma}
\newskip\humongous \humongous=0pt plus 1000pt minus 1000pt
\def\caja{\mathsurround=0pt}
\def\eqalign#1{\,\vcenter{\openup2\jot \caja
        \ialign{\strut \hfil$\displaystyle{##}$&$
        \displaystyle{{}##}$\hfil\crcr#1\crcr}}\,}
\newif\ifdtup

\def\a{\alpha}

\def\ha{{1\over 2}}

\documentstyle[12pt]{article}

\makeatletter                    
\@addtoreset{equation}{section}  
\makeatother                     

\begin{document}

\title{Quantum Brownian Motion in a Bath of Parametric Oscillators:
A model for system-field interactions}

\author{B. L. Hu\thanks{ Email: hu@umdhep.umd.edu}\\
{\small Department of Physics, University of Maryland,
College Park, MD 20742, USA} \\ A. Matacz\thanks{ Email:
amatacz@physics.adelaide.edu.au}\\
{\small Department of Physics, University of Adelaide, 5005, Australia}}
\date{3 December 1993}
\maketitle
\vskip-\baselineskip
\centerline{(umdpp 93-210) }

\begin{abstract}

The quantum Brownian motion paradigm provides a unified framework
where one can see the interconnection of some basic quantum statistical
processes like decoherence, dissipation, particle creation,
noise and fluctuation.
The present paper continues the investigation into these issues
begun in two earlier papers by Hu, Paz and Zhang on the quantum Brownian
motion in a general environment via the influence functional formalism.
Here, the Brownian particle is coupled linearly to a bath of the most general
time dependent quadratic oscillators.
This bath of parametric oscillators mimics a scalar field, while
the motion of the Brownian particle modeled by a single
oscillator could be used to depict the behavior of a particle detector,
a quantum field mode or the scale factor of the universe.
An important result of this paper is the derivation of the influence functional
encompassing the noise and dissipation kernels in terms of the Bogolubov
coefficients, thus
setting the stage for the influence functional formalism treatment
of problems in quantum field theory in curved spacetime.
This method enables one to trace the source of statistical processes
like decoherence and dissipation to vacuum fluctuations and particle creation,
and in turn impart a statistical mechanical interpretation of quantum field
processes. With this result we discuss the statistical mechanical origin of
quantum noise and thermal radiance from black holes and from
uniformly-accelerated observers in Minkowski space as well as from the
de Sitter universe discovered by Hawking, Unruh and Gibbons-Hawking.
We also derive the exact evolution operator
and master equation for the reduced density matrix of the system interacting
with a parametric oscillator bath in an initial squeezed thermal state.
These results are useful for decoherence and backreaction studies for
systems and processes of interest in semiclassical cosmology
and gravity.  Our model and results are also expected to be useful
for related problems in quantum optics.

\end{abstract}
pacs{05.40.+j,03.65.Sq,98.80.Cq,97.60.Lf}
\newpage

\section{Introduction}

In two earlier papers, called Paper I, II henceforth \cite{HPZ1,HPZ2},
Paz, Zhang and one of us
began a systematic study of the celebrated problem
of quantum Brownian motion (QBM) in a general environment using
the Feynman-Vernon influence functional (IF) formalism
\cite{FeyVer,CalLeg83,Gra}. The special features
associated with a nonohmic bath, or ohmic bath at low temperatures are
the appearance of colored noise and nonlocal dissipation. The motivation for
this study was amply explained there. What prompted them to this undertaking
was the belief that a correct and deepened understanding of many interesting
quantum statistical processes in the early universe and black holes
\cite{HuWaseda} requires an extension
of the existing framework of quantum field theory in curved spacetime
\cite{BirDav} to statistical and stochastic fields in the framework of quantum
open systems \cite{qos} represented by the QBM \cite{Zhang}.
This viewpoint and methodology have indeed been
applied to the analysis of some basic issues in quantum cosmology
\cite{HuErice,HuTsukuba,Sinha,PazSin,HPS,decQC},
effective field theory \cite{HuPhysica,HuBanff},
and the foundation of quantum mechanics, such as the uncertainty principle
\cite{HuZhaUncer,AndHal} and, most significantly, decoherence
\cite{envdec,conhis,GelHar1,GelHar2,CalHuDCH}
in the quantum to classical transition problem. (See the recent reviews of
\cite{ZurekPTP,HarLH,Omnes} and references therein and in Papers I, II
for the standard literature on this topic). QBM is one of the two
major paradigms of non-equilibrium statistical mechanics (the other being
Boltzmann's kinetic theory) which  is also amenable to detailed analysis.
The study of many problems mentioned above
which have nonlinear and nonlocal characteristics typical of quantum
processes in gravitation and cosmology necessitates a closer
scrutiny of this model beyond the ordinary limited conditions.

As stated in the Introductions of Papers I and II,
in order to make it useful for
addressing issues in semiclassical gravity and quantum cosmology,
a theory of quantum open systems has to be developed for quantum fields
in curved spacetime.
Noticeable effort has been put into this direction. Hu, Paz and Zhang
\cite{qsf1} constructed a stochastic field theory based on the QBM model
and described how thermal field theory can be obtained as the equilibrium
limit. As a tool for the
study of the quantum origin of noise, fluctuations and structure formation
in cosmology, they \cite{HuBelgium} have extended the result of
Paper II to quantum fields in Minkowski, Robertson-Walker and
de Sitter spacetime.
The nature and origin of quantum noise from particle-field interaction
were discussed in \cite{HMLA} where a statistical field-theoretical
derivation of thermal radiance in the Hawking \cite{Haw75,GibHaw}
and Unruh effects \cite{Unr} were given.
For semiclassical gravity Kuo and Ford \cite{KuoFor} have studied the
fluctuations of quantum fields on the Einstein equations.
Calzetta and Hu \cite{CalHuSG}, and the present authors \cite{HM3}
have analyzed the nature of noise, fluctuations, particle creation and
backreaction for quantum fields in cosmological spacetimes
and proposed an Einstein-Langevin equation as the centerpiece of
a generalized theory of semiclassical gravity. For quantum cosmology,
Sinha and Hu \cite{SinHu} had used the coarse-grained \cite{cgea}
Schwinger-Keldysh effective action \cite{ctp} to analyze the validity of the
minisuperspace approximation
in quantum cosmology. Paz and Sinha \cite{PazSin} had used the
influence functional method to discuss the transition from quantum
to semiclassical gravity, and Calzetta and Hu \cite{disQG} have
studied dissipation
problem in quantum cosmology. However, except for the few cases mentioned
above,
none of these earlier work made use of
the master or Langevin equation approach characteristic of the QBM study,
which is necessary to probe into the noise, fluctuation
\cite{CalHuSG,HM3}, instability and phase transition \cite{HuBanff}
aspects of quantum fields and spacetime.

The present paper is an intermediate step in that direction. It is a
generalization of Papers I and II in that the oscillators which make up the
system and bath are now the most general time-dependent quadratic oscillators.
This bath of parametric
oscillators (as the number of modes goes to infinity) is identical to a
scalar field, while the motion of the Brownian particle modeled by a single
oscillator could be used to depict the behavior of a particle detector
(with zero spring constant, as in e.g., \cite{Unr}), the scale factor
of the universe,
(with a negative kinetic energy term, as is seen in Eq.(2.2) of \cite{SinHu})
or the homogeneous or inhomogeneous
(density fluctuation) modes of the inflaton field in an early universe
\cite{inflation,stoinf,galforinf,GutPi,CorBru,decinf,Mat}. Indeed the results
obtained here
can be taken over directly for the description of
scalar fields in cosmological spacetimes, as our examples will demonstrate.
Parametric amplification of the bath oscillator quanta gives rise to
particle creation, as was pointed out by Parker and Zel'dovich \cite{cpc},
which can be depicted by the Bogolubov transformation between the creation
and annihilation operators of the Fock spaces defined at different times.
The averaged effect of the bath on the system is described by
the influence functional, which, in the statistical field-theory context
measures the backreaction of quantum processes associated with the field
like particle creation on the dynamics of the background spacetime
\cite{cosbkr,dissip}.
There are two components in the influence functional, a noise kernel and a
dissipation kernel. The noise kernel governs the
decoherence process and also limits the degree of attainment of classicality
\cite{GelHar2}. It also depicts the effect of fluctuations (in particle number)
\cite{CalHuSG}. The dissipation kernel which appears in the effective
equation of motion depicts the effect of particle creation on the
dynamics of the system. The QBM paradigm thus provides a unified
framework where one can see the interconnection of the basic quantum
statistical processes like decoherence,
dissipation, particle creation, noise and fluctuation. The neccesity of
analyzing these processes on the same footing was emphasized earlier
in \cite{HuTsukuba}.

An important result of this paper is the derivation of the influence functional
and thus the noise and dissipation kernels in terms of the Bogolubov
coefficients. This enables one to trace the source of statistical processes
like decoherence and dissipation to vacuum fluctuations and particle creation,
and in turn impart a statistical mechanical interpretation of quantum field
processes. With this we discuss the well-known results by Unruh \cite{Unr},
Hawking \cite{Haw75} and Gibbons-Hawking \cite{GibHaw} on thermal radiance
from uniformly-accelerated observers \cite{Ang}, black holes and for
comoving observers in de Sitter spacetime.
{}From the explicit
form of the noise and dissipation kernels we derived, one can see clearly the
interplay of different factors which contribute to making the spectrum of
particle creation in these situations thermal, and, more interestingly,
what makes them nonthermal, as in the more general non-equilibrium situations.
This is where the capability of the statistical field-theoretical
interpretation
supercedes the geometric interpretation (in terms of event horizons). We will
discuss the implications of this point later.

Although we have used examples from quantum and semiclassical cosmology to
illustrate
the physical relevance of the QBM model with parametric bath, the
range of applicability of this model goes beyond. An important area
where parametric amplification plays a central
role is in quantum optics. Here the properties of baths prepared in
squeezed initial states (rigged reservoirs) are of interest
\cite{milburn,gilson}. Squeezed baths are capable of processing optical signals
(attenuation or amplification) while retaining their quantum features.
It has also been shown that an appropriately
squeezed bath is capable of greatly increasing the decoherence timescale
\cite{sqth}. The description of these processes is based on the
quantum optical master equation generalised to include squeezing in the
initial state. It is an approximate equation derived under the
rotating wave, Born and Markov approximations. Since our formalism is
exact it is capable of a more accurate description of non-equilibrium
quantum statistical processes in quantum
optics. It also allows for the squeezing to be generated dynamically
rather than imposed as an initial condition.

The effect of the bath on the system is studied here, as in the
previous two QBM papers, by means of the influence functional formalism.
We will derive exactly the evolution operator for the reduced density matrix,
the
influence functional, and the master equation for a time-dependent system and
bath, using a slightly different method and language from Paper I.
We adopt the language of squeeze and rotation operators \cite{sqst,GriSid,HKM}
for describing the evolution of the system. In Sec. 2
we define the model and mention its relevance in problems in quantum optics,
quantum and semiclassical cosmology, and quantum field theory in curved
spacetimes. We then
derive an analytic expression for the influence functional of a system
linearly coupled to a bath of parametric oscillators in terms of the
Bogolubov coefficients. In Sec. 3 we derive the exact evolution operator for
the
reduced density matrix and
adopt the simpler method introduced by Paz \cite{PazSpain} and used in
\cite{qsf1} for the derivation of the master equation. We consider the
general case when the bath is initially in a squeezed thermal state,
which includes the common cases of a thermal state and a squeezed vacuum.
We indicate how it is different from the model with a bath of time independent
oscillators.
The diffusion coefficients of this  equation can be analyzed for
decoherence studies, as is done in Papers I, Refs. \cite{HuBelgium}
and \cite{envdec}.
The relation of decoherence and particle creation was also discussed
in the field theory context by Calzetta and Mazzitelli \cite{CalMaz}
and in the quantum cosmology context by Paz and Sinha \cite{PazSin}.
Here we aim not at the decoherence or the dissipation processes,
but focus on the definition and nature of noise associated with
quantum fields and use them to depict some well-known processes such as the
Hawking effect in gravitation and cosmology.

In Sec. 4 we give a few simple examples of a system interacting with a bath of
parametric oscillator, first treating the case with constant frequency, but
with
an initial squeezed thermal state and then the case of inverted oscillators
which can be used to model amplifiers in quantum optics and electronics
\cite{gilson}.
In Sec. 5, we derive the noise kernels for four cases: the accelerated
observer, a two-dimensional black hole and a massless, conformally and
minimally coupled scalar field in the de Sitter universe. In the de Sitter
universe case the parametric oscillator bath can serve as a relatively simple
model of the environment for homogenous and inhomogenous (density fluctuation)
modes of the inflaton field in the early universe.
We show the factors entering into the determination of the spectrum,
and indicate how one can
understand the Hawking and Unruh effects in a purely statistical-mechanical
sense without recourse to geometric notions (like the event horizon).
We will discuss the fluctuation-dissipation relation approach
\cite{fdr,Sciama,Mottola} to understanding backreaction in
semiclassical gravity in later work \cite{HuPhysica,HuSin94,CalHuSG,HM3}.
In Sec. 6 we summarize our results and suggest further problems in cosmology
and gravitation where our results can be usefully applied.
The details of derivation in Sec. 2 are recorded in the Appendices.

\section{Influence Functional}
Our system, the Brownian particle, is modeled by
a parametric oscillator with mass $M(s)$, cross term $B(s)$
and natural (bare) frequency $\Omega(s)$. The environment (bath) is also
modeled by a set of parametric oscillators with mass
$m_n(s)$, cross term $b_n(s)$  and natural frequency $\omega_n(s)$.
The system is  coupled to the bath through an arbitrary function $F(x)$
of the system variable and linear in the bath variables $q_n$ with coupling
strength $c_n(s)$ in each oscillator. The action of the combined
system $+$ environment is

$$
{\eqalign{S[x,{\bf q}]
& =S[x]+S_E[{\bf q}]+S_{int}[x,{\bf q}] \cr
& =\int\limits_0^tds \Biggl[ {1\over 2}
  M(s)\Bigl( \dot x^2+B(s)x\dot{x}-\Omega^2(s)x^2\Bigr) \cr
& + \sum_n\Bigl\{  {1\over 2}m_n(s)
  \Bigl(\dot q_n^2+b_n(s)q_n\dot{q}_n- \omega^2_n(s)q_n^2\Bigr) \Bigr\}
  + \sum_n\Bigl(-c_n(s)F(x)q_n\Bigr)\Biggr]\cr}}\eqno(2.1)
$$

\noindent where $x$ and $q_n$ are the coordinates of the particle and the
oscillators. The bare frequency $\Omega$ is different from the physical
frequency $\Omega_p$ due to its interaction with the bath, which depends
on the cutoff frequency. We will discuss this point in more detail in
Sec. 4.1. For problems discussed here
we are interested in how the environment affects the system in
some averaged way.  The quantity containing this information is the reduced
density matrix of the system obtained from the full density operator of the
system $+$ environment by tracing out the environmental degrees of freedom. The
evolution operator is responsible for the time evolution of the reduced density
matrix.
The equation of motion governing this reduced density matrix is the master
equation.
Our central task is to derive the evolution operator and the master equation
for the Brownian particle in a general environment.

We will briefly review here the Feynman-Vernon
influence functional method
for deriving the evolution operator. Readers who are familiar with it can skip
this subsection. The method provides an easy way to obtain a functional
representation for the evolution operator for the reduced density matrix
$ {\cal J}_r $. Let us start first with the evolution operator for the full
density
matrix $ {\cal J} $ defined by
$$
\hat\rho(t)={\cal J}(t,t_i)\hat\rho(t_i). \eqno(2.2)
$$

As the full density matrix $\hat\rho$ evolves unitarily under the
action of (2.1), the evolution operator ${\cal J}$ has a simple path integral
representation. In the position basis, the matrix elements of the evolution
operator are given by

$$\eqalign{{\cal J}(x,{\bf q},x',{\bf q}',t~&|~x_i,{\bf q}_i,x'_i,{\bf
q}'_i,t_i)
= {\cal K}(x,{\bf q},t~|~x_i,{\bf q}_i, t_i)
  {\cal K}^*(x',{\bf q}',t~|~x'_i,{\bf q}'_i, t_i)\cr
  &= \int\limits_{x_i}^{x}Dx\int\limits_{{\bf q}_i}^{{\bf q}}D{\bf
q}~\exp{i\over\hbar}S[x,{\bf q}]
     \int\limits_{x'_i}^{x'}Dx'\int\limits_{{\bf q}'_i}^{{\bf q}'}D{\bf
q}'~\exp{-{i\over\hbar}}
     S[x',{\bf q}']}\eqno(2.3)
$$

\noindent where the operator ${\cal K}$ is the evolution operator for the wave
functions. In the second equation, the path integrals are over all histories
compatible with the boundary conditions. We have used  ${\bf q}$ to represent
the
full set of environmental
coordinates $q_n$ and the subscript $i$ to denote the initial variables.

The reduced density matrix is defined as

$$
\rho_r(x,x')
=\int\limits_{-\infty}^{+\infty}dq\int\limits_{-\infty}^{+\infty}
  dq'\rho(x,{\bf q};x',{\bf q}')\delta({\bf q}-{\bf q}')\eqno(2.4)
$$

\noindent and is propagated in time by the evolution operator ${\cal J}_r$
$$
\rho_r(x,x',t)
=\int\limits_{-\infty}^{+\infty}dx_i\int\limits_{-\infty}^{+\infty}dx'_i~
 {\cal J}_r(x,x',t~|~x_i,x'_i,t_i)~\rho_r(x_i,x'_i,t_i~).\eqno(2.5)
$$
By using the functional representation of the full density matrix evolution
operator given in (2.3), we can also represent ${\cal J}_r$ in path integral
form.
In general, the expression is very complicated since the evolution operator
${\cal J}_r$ depends on the initial state. If we assume that at a given
time $t=t_i$ the system and the environment are uncorrelated
$$
\hat\rho(t=t_i)=\hat\rho_s(t_i)\times\hat\rho_b(t_i),\eqno(2.6)
$$
then the evolution operator for the reduced density matrix does not depend
on the initial state of the system and can be written \cite{FeyVer} as
$$
\eqalign{
{\cal J}_r(x_f,x'_f,t~|~x_i,x'_i,t_i)
& =\int\limits_{x_i}^{x_f}Dx
   \int\limits_{x'_i}^{x'_f}Dx'~
   \exp{i\over\hbar}\Bigl\{S[x]-S[x']\Bigr\}~{\cal F}[x,x']  \cr
& =\int\limits_{x_i}^{x_f}Dx
   \int\limits_{x'_i}^{x'_f}Dx'~
   \exp{i\over\hbar} {\cal A}[x,x']  \cr }
                                                       \eqno(2.7)
$$
where the subscript $f$ denotes final variables, and ${\cal A}[x,x']$ is the
effective action for the open quantum system.
The factor ${\cal F}[x,x']$,  called the `influence functional', is defined as
$$
\eqalign{ {\cal F}[x,x']= &\int\limits_{-\infty}^{+\infty}d{\bf q}_f
 \int\limits_{-\infty}^{+\infty}d{\bf q}_i
 \int\limits_{-\infty}^{+\infty}d{\bf q}'_i
 \int\limits_{{\bf q}_i}^{{\bf q}_f}D{\bf q}
  \int\limits_{{\bf q}'_i}^{{\bf q}_f}D{\bf q}' \cr
& \times \exp{i\over\hbar}\Bigl\{
  S_b[{\bf q}]+S_{int}[x,{\bf q}]-S_b[{\bf q}']-S_{int}[x',{\bf q}'] \Bigr\}
  \rho_b({\bf q}_i,{\bf q}'_i,t_i) \cr
& = \exp{i\over\hbar} \delta {\cal A}[x,x'] \cr}                \eqno(2.8)
$$
where $\delta {\cal A}[x,x']$ is the influence action.  Thus $ {\cal A}[x,x'] =
S[x]-S[x'] + \delta {\cal A}[x,x']$.

It is not difficult to show that (2.8) has the representation independent
form
$$
{\cal F}[x,x']=Tr[\hat{U}(t,t_i)\hat{\rho}_b (t_i)\hat{U}'^{\dag}(t,t_i)]
\eqno(2.9)
$$
where $\hat{U}(t)$ and $\hat{U}'(t)$ are the quantum propagators for the
actions \\ $S_E[{\bf q}]+S_{int}[x(s),{\bf q}]$ and $S_E[{\bf
q}]+S_{int}[x'(s),{\bf q}]$ and
$x(s)$ and $x'(s)$ are treated as time dependent classical forcing terms.
We have found this form to be much more convenient for deriving the influence
functional.

It is obvious from its definition that if the interaction term is zero, the
influence functional is equal to unity and the influence action is zero.
In general, the influence functional
is a highly non--local object. Not only does it depend on the time history,
but --and this is the more important property-- it also
irreducibly mixes the two sets
of histories in the path integral of (2.7). Note that the histories
$ x $ and $ x' $ could be interpreted as moving
forward and backward in time respectively.
Viewed in this way, one can see the similarity of the influence functional
\cite{FeyVer} and the generating functional in the closed-time-path
(CTP or Schwinger-Keldysh) integral formalism \cite{ctp}.
The Feynman rules derived in the CTP method are very useful for computing
the IF. We shall treat the field theoretic problems in later papers.

In those cases where the initial decoupling
condition $(2.6)$ is satisfied, the influence functional
depends only on the initial state of the environment. The influence functional
method can be extended to more general
conditions, such as thermal equilibrium between the system and the environment
\cite{HakAmb}, or correlated initial states \cite{Gra,CalLeg83}.

We now proceed to derive the influence functional for the model (2.1).
{}From its definition it is clear that the influence functional is independent
on the choice of system but only on the coupling of the system to
the environment. Since our method is quite
general we have been able to include, in Appendix A,
the influence functional for
the most general coupling linear in the bath variable.
However in the body of the paper we only
consider the position-position coupling in (2.1). For the case of a
squeezed thermal initial state (to be defined later) we find that
for the model (2.1) the influence functional has the form
$$
\eqalign{
{\cal F}[x,x']=\exp\biggl\{
& -{i\over\hbar}\int\limits_{t_i}^tds\int\limits_{t_i}^{s}ds'
   \Bigl[F(x(s))-F(x'(s))\Bigr]\mu(s,s')
   \Bigl[F(x(s'))+F(x'(s'))\Bigr] \cr
& -{1\over\hbar}\int\limits_{t_i}^tds\int\limits_{t_i}^{s}ds'
   \Bigl[F(x(s))-F(x'(s))\Bigr]\nu(s,s')
   \Bigl[F(x(s'))-F(x'(s'))\Bigr]\biggl\}. \cr}
                                                        \eqno(2.10)
$$
The functions $\mu(s,s')$ and $\nu(s,s')$ contain the effects of the
environment on the system. They are known respectively as the dissipation
and noise kernels. The reason for these names becomes clear in the
semi-classical regime of the open system generated by (2.10).

To find the appropriate semiclassical limit
of this open quantum system we must find an action which generates
the same influence functional as (2.10). Consider the action
$$
S[a(s)]=\int_{t_i}^{t}ds\Bigl(L(x,\dot{x},s)
+F(x)\xi(s)\Bigr)
\eqno(2.11)
$$
where $\xi(s)$ is a gaussian stochastic force with a non-zero mean. This system
generates the influence functional
$$
{\cal F}[\Sigma,\Delta]=\biggl\langle\exp\left[\frac{i}{\hbar}\int_{t_i}^{t_f}
\xi(s)\Delta(s)ds\right]\biggr\rangle
\eqno(2.12)
$$
where $\Sigma$ and $\Delta$ are given by
$$
\Sigma(s)=\frac{1}{2}\Bigl(F(x(s))+F(x'(s))\Bigr),\;\;\;
\Delta(s)=F(x(s))-F(x'(s))
\eqno(2.13)
$$
and the average is understood as a functional integral over $\xi(s)$ multiplied
by a normalised gaussian probability density functional ${\cal
P}[\xi(s);\Sigma(s)]$.
The probablility density functional is a functional of $\Sigma(s)$ if we allow
the statistical properties of $\xi$ to depend on the system history.
The avergaing can be performed to give \cite{HM3}
$$
{\cal F}[x,x']=\exp\biggl\{
  {i\over\hbar}\int\limits_{t_i}^tds\Delta(s)\langle \xi(s)\rangle
-{1\over\hbar^2}\int\limits_{t_i}^tds\int\limits_{t_i}^{s}ds'
   \Delta(s)\Delta(s')C_2(s,s')\biggl\}
\eqno(2.14)
$$
where $C_2(s,s')$ is the second cumulant of the force $\xi$.
The equation of motion generated by the action (2.11) is
$$
\frac{\partial L}{\partial x}-\frac{d}{dt}\frac{\partial L}{\partial\dot{x}}
+\frac{\partial F(x)}{\partial x}\langle \xi(t)\rangle=-\frac{\partial
F(x)}{\partial x}\langle\bar{\xi}(t)\rangle
\eqno(2.15)
$$
where $\bar{\xi}(t)$ is a zero mean gaussian stochastic force with
$\langle\bar{\xi}(t)\bar{\xi}(t')\rangle=C_2(s,s')$.
Now by comparing (2.14) and (2.10) we see that
$$
\langle\xi(s)\rangle\equiv -2\int_{t_i}^s ds'\mu(s,s')\Sigma(s'),\;\;
C_2(s,s')\equiv\hbar\nu(s,s').
\eqno(2.16)
$$
Therefore the semiclassical equation for the system described by the
influence functional (2.10) is
$$
\frac{\partial L}{\partial x}-\frac{d}{dt}\frac{\partial L}{\partial\dot{x}}
-2\frac{\partial F(x)}{\partial x} \int_{t_i}^t
\mu(t,s)F(x(s))ds=-\frac{\partial F(x)}{\partial x}\bar{\xi}(t)
\eqno(2.17)
$$
where $\langle \bar{\xi}(t)\bar{\xi}(t')\rangle=\hbar\nu(t,t')$.
Under special circumstances $\mu$ tends to the derivative of a delta function
which generates local dissipation. More generally
we see that in the semiclassical limit $\mu$ generates non-local
dissipation while $\hbar\nu$ is the correlator of a zero mean gaussian
stochastic force.

We find that the dissipation and noise kernels take the form
$$
\eqalign{
\mu(s,s')=\frac{i}{2}\int_0^{\infty}d\omega I(\omega,s,s')&
\biggl[\{\a_{\o}(s)+\b_{\o}(s)\}^*
\{\a_{\o}(s')+\b_{\o}(s')\} \cr
&-\{\a_{\o}(s)+\b_{\o}(s)\}
\{\a_{\o}(s')+\b_{\o}(s')\}^*\biggr] \cr} \eqno(2.18)
$$
$$
\eqalign{
\nu(s,s')= \frac{1}{2}\int_0^{\infty}&d\omega I(\omega,s,s')
\coth\left(\frac{\hbar\omega(t_i)}{2k_BT}\right) \cr
& \times\biggl[\cosh 2r(\o)\{\a_{\o}(s)+\b_{\o}(s)\}^*
  \{\a_{\o}(s')+\b_{\o}(s')\} \cr
& +\cosh 2r(\o)\{\a_{\o}(s)+\b_{\o}(s)\}
  \{\a_{\o}(s')+\b_{\o}(s')\}^* \cr
&-\sinh 2r(\o) e^{-2i\phi(\o)}\{\a_{\o}(s)+\b_{\o}(s)\}^*
  \{\a_{\o}(s')+\b_{\o}(s')\}^*  \cr
& -\sinh 2r(\o) e^{2i\phi(\o)}\{\a_{\o}(s)+\b_{\o}(s)\}
  \{\a_{\o}(s')+\b_{\o}(s')\} \biggr]. \cr} \eqno(2.19)
$$
The quantities in these kernels describe three different properties of
the environment.

A) The $\a$ and $\b$, known as the Bogolubov coefficients, are complex
numbers that contain all the information about the quantum dynamics of the bath
parametric oscillators. They are derived from two coupled first order equations
$$
\eqalign{
&\dot{\a}_n=-if^*_n\b_n-ih_n\a_n \cr
&\dot{\b}_n=ih_n\b+if_n\a \cr} \eqno(2.20)
$$
where the time dependent coefficients are given by
$$
\eqalign{
&f_n(t)=\frac{1}{2}\left(\frac{m_n(t)\omega_n^2(t)}{\kappa_n}
+\frac{m_n(t)b_n^2(t)}{4\kappa_n}-\frac{\kappa_n}{m_n(t)}+ib_n(t)\right) \cr
&h_n(t)=\frac{1}{2}\left(\frac{\kappa_n}{m_n(t)}+\frac{m_n(t)
\omega_n^2(t)}{\kappa_n}
+\frac{m_n(t)b_n^2(t)}{4\kappa_n}\right). \cr}
\eqno(2.21)
$$
These equations are a  by product of finding the quantum propagator
for a parametric oscillator which is done in appendix B.
We will usually choose $\kappa_n$ (defined by (A.6)) so that $f(t_i)=0$.
Thus if $b_n=0$ we will usually have $\kappa_n=m_n(t_i)\o_n(t_i)$.
Eq's (2.21) must satisfy the initial conditions $\alpha(t_i)=1, \beta(t_i)=0$.
Note that
the mode label $\o$ in the kernels is equivalent to $n$ in the continuous
limit.

If we assume $b=0$ and $m=1$ we can show using (2.20) that
$$
\ddot{X}_n+\omega^2_n(t)X_n=0
\eqno(2.22)
$$
where $X_n(t)=\alpha_{n}(t)+\beta_{n}(t)$. The solution of (2.22) must satisfy
$X_n(t_i)=1$. In this case the noise and dissipation kernels become
$$
\mu(s,s')=\frac{i}{2}\int_0^{\infty}d\omega I(\omega,s,s')
\biggl[X_{\o}^*(s)X_{\o}(s')-X_{\o}(s)X_{\o}^*(s')\biggr]
\eqno(2.23)
$$
$$
\eqalign{
\nu(s,s')= \frac{1}{2}\int_0^{\infty}&d\omega I(\omega,s,s')
\coth\left(\frac{\hbar\omega(t_i)}{2k_BT}\right)
\biggl[\cosh 2r(\o)\Bigl[X_{\o}^*(s)X_{\o}(s')+X_{\o}(s)X_{\o}^*(s')\Bigr] \cr
&-\sinh 2r(\o)\Bigl[ e^{-2i\phi(\o)}X_{\o}^*(s)X_{\o}^*(s')+
e^{2i\phi(\o)}X_{\o}(s)X_{\o}(s')\Bigr]
\biggr]. \cr} \eqno(2.24)
$$
Note that we can always write
$$
X_n(t)=C_{n}(t)-i\o_n(t_i)S_{n}(t)
\eqno(2.25)
$$
where $C_n$ and $S_n$ are subject to the boundary conditions
$C_n(t_i)=\dot{S}_n(t_i)=1$ and
$S_n(t_i)=\dot{C}_n(t_i)=0$.
If the kernels are written in this notation we can show
that for a  thermal initial state
(2.17) reduces to the classical Langevin equation in the high temperature limit
\cite{HabKan}.\\

\noindent B) The spectral density, $I(\o,s,s')$ defined formally by
$$
I(\omega,s,s')=\sum_n\delta(\omega-\omega_n)\frac{c_n(s)c_n(s')}{2\kappa_n}
\eqno(2.26)
$$
is obtained in the continuum limit. It
contains information about the environmental mode density and coupling
strength as a function of frequency.
Different environments are classified according to the
functional form of the spectral density $I(\o)$.  On physical grounds, one
expects the spectral density to go to zero for very high frequencies.  Let
us introduce a certain cutoff frequency $\Lambda$  (a property of the
environment) such that $I(\o)\rightarrow 0$ for $\omega>\Lambda$.
The environment is
classified as ohmic \cite{CalLeg83,Gra}
if in the physical range of frequencies
($\omega<\Lambda$) the spectral density is such that
$I(\o)\sim\omega$, as supra-ohmic if $I(\o)\sim\omega^n , n>1$
or as sub-ohmic if $n<1$. The most studied ohmic case corresponds to an
environment which induces a dissipative force linear in the velocity of the
system. We will show this in section 4.1.\\

\noindent C) The initial state of the bath is a squeezed thermal state.
It has the form
$$
\hat{\rho}_b(t_i)=\prod_n\hat{S}_n(r(n),\phi(n))\hat{\rho}_{th}
\hat{S}_n^{\dag}(r(n),\phi(n))
\eqno(2.27)
$$
where $\hat{\rho}_{th}$ is a thermal
density matrix of temperature $T$ defined by (A.18)  and $\hat{S}(r,\phi)$
is a squeeze operator defined by (B.12).
Since a squeezed thermal state is still gaussian it is a tractable
generalisation
of the usual thermal initial state that is of interest in quantum optics
\cite{sqth}.
For the case of zero temperature we have a squeezed
vacuum initial state.

Physically the term squeezing arises because the
phase space noise distribution of a squeezed vacuum is an ellipse squeezed
from a circle (coherent state) by $r$ and rotated by angle $\phi$ with respect
to the $x$ and $p$ axes. Thus, for the squeezed vacuum \cite{sqst}
$$
\eqalign{
&\langle q^2\rangle=\frac{1}{2\kappa}[\cosh 2r-\sinh 2r\cos 2\phi]\cr
&\langle p^2\rangle=\frac{1}{2\kappa}[\cosh 2r+\sinh 2r\cos 2\phi]. \cr}
\eqno(2.28)
$$
Note that the dissipation kernel is independent
of the bath initial state.

For the case of no initial squeezing we see that the noise and
dissipation kernels  are built out of symmetric and anti-symmetric combinations
of
identical Bogolubov factors. Thus the two kernels are intimately linked.
For the case when the bath is a standard harmonic oscillator this
interelationship can be written as a generalised fluctuation-dissipation
relation \cite{HPZ2}.

\section{Evolution Operator and Master Equation}
In this section our goal is to calculate the evolution operator for
the reduced density matrix and the master equation. The master equation is the
evolution equation for the reduced density matrix.
It provides a transparent
means for sifting out the different physical processes caused by the bath on
the system. First  we must calculate the evolution operator $ \rho_r$ in
(2.7), which contains all the dynamical information about the open system.
{}From this point on we shall put $F(x)=x$.

The influence functional (2.10) and the corresponding influence action (2.8)
can be written in a compact way
$$
\eqalign{
{\cal A}[x,x']&= S[x] - S[x'] + \delta {\cal A}(x,x'),\cr
\delta {\cal A}[x,x']& = -~2\int_{t_i}^tds\int_{t_i}^{s}ds'~
  \Delta(s)\mu(s,s')\Sigma(s')~
 +i\int_{t_i}^t ds\int_{t_i}^{s}ds'
  \Delta(s)\nu(s,s')\Delta(s') }
\eqno(3.1)
$$
$$
\eqalign{
S[x]-S[x']&=\int_{t_i}^tds \{ M(s)\dot{\Sigma}(s)\dot{\Delta}(s)
+\ha M(s)B(s) [\Sigma(s)\dot{\Delta}(s)+\Delta(s)\dot{\Sigma}(s)] \cr
& -M(s)\Omega^2(s)\Sigma(s)\Delta(s) \} }
\eqno(3.2)
$$
with the use of the `center of mass' and `relative'
coordinates defined earlier in (2.13).

As pointed out by many authors \cite{FeyVer,CalLeg83,Gra}, and in Sec. 2, the
real and imaginary parts of ${\cal A}[x,x']$ can be interpreted \cite{FeyVer}
as being responsible for dissipation and noise respectively.
The imaginary part of (3.1)
is determined by $\nu(s)$, the noise (or fluctuation)
kernel. The name becomes apparent when we realize that this term can be
interpreted as coming from the interaction
between the system and a stochastic force $\xi$ that is linearly coupled to the
system and has a probability density given by
${\cal P}[\xi]=exp\{-\xi(\hbar\nu)^{-1}\xi\}$.
On the other hand, the kernel $\mu(s)$ in $(3.1)$
is known as the dissipation kernel.
The motivation for the name comes from the fact \cite{Zhang}
that it introduces a modification in the real saddle point
trajectories of the path integral in (2.7). Strictly speaking
only the non-symmetric part of the $\mu$ kernel should be associated with
dissipation.
Thus, all the symmetric part can be absorbed in a non-local potential
(that does not contribute to the mixing of the
$ x $ and $ x' $ histories). There is no such symmetric part
in the $\mu$--kernel of
our problem although it does appear in other cases \cite{HPZ2}.

\subsection{ Evolution Operator}

The evolution operator given in equation $(2.7)$ generates a
non--Markovian dynamics
since it fails in general to satisfy the relation
$${\cal J}_r(t_2,t_i)={\cal J}_r(t_2,t_1)~{\cal J}_r(t_1,t_i)$$
\noindent for the reason that
the operator ${\cal J}_r(t_2,t_1)$ depends on the state of the system at
time $t_1$, unless that time is the one for which the system and the
environment were decoupled. The non--Markovian behavior is, in fact,
a direct consequence of the non--locality of the influence functional.

Our task is to compute the evolution operator
$$
{\cal J}_r(\S_f,\D_f,t~|~\S_i,\D_i,t_i)
  =\int_{\S_i}^{\S_f}D\S
   \int_{\D_i}^{\D_f}D\D~
   \exp\left[\frac{i}{\hbar}{\cal A}[\S(s),\D(s)]\right].
\eqno(3.3)
$$
Let us schematically describe how to compute the path integral in (3.3).
We start by reparametrizing the paths, writing
$$
\eqalign{
& \S(s)=x_+(s)+\S_{cl}(s) \cr
& \D(s)=x_-(s)+\D_{cl}(s) \cr }
             \eqno(3.4)
$$
\noindent where the ``classical paths''
${\pmatrix{\Sigma\cr\Delta\cr}}_{cl}$
are solutions to the equations of motion derived from the real part of
${\cal A}[\Sigma,\Delta]$, and $x_\pm$ are the deviations from the classical
paths.
The equations governing these functions are

$$
\eqalign{
& \ddot\Sigma_{cl}(s)+\frac{\dot{M}(s)}{M(s)}\dot{\Sigma}_{cl}(s)+
\left(\Omega^2(s)+\frac{\dot{B}(s)}{2}+\frac{\dot{M}(s)B(s)}{2M(s)}\right)
\Sigma_{cl}(s)+
 \frac{2}{M(s)}\int\limits_{t_i}^sds'\mu(s,s')\Sigma_{cl}(s')=0 \cr
& \Sigma_{cl}(t_i)=\Sigma_i, \qquad {and}\qquad
  \Sigma_{cl}(t)=\Sigma_f \cr }
                                                        \eqno(3.5)
$$
\noindent and
$$
\eqalign{
&  \ddot\Delta_{cl}(s)+\frac{\dot{M}(s)}{M(s)}\dot{\Delta}_{cl}(s)+
\left(\Omega^2(s)+\frac{\dot{B}(s)}{2}+\frac{\dot{M}(s)B(s)}{2M(s)}\right)
\Delta_{cl}(s)
+\frac{2}{M(s)}\int\limits_s^tds'\mu(s',s)\Delta_{cl}(s')=0 \cr
& \Delta_{cl}(t_i)=\Delta_i, \qquad {and}\qquad
  \Delta_{cl}(t)=\Delta_f.  \cr }
                                                        \eqno(3.6)
$$
After the path-reparametrization, (3.3) can be rewritten as
$$
{\cal J}_r(\S_f,\D_f,t~|~\S_i,\D_i,t_i)
=Z(t,t_i)\exp\left[\frac{i}{\hbar}{\cal A}[\S_{cl}(s),\D_{cl}(s)]\right]
\eqno(3.7)
$$
where
$$
\eqalign{Z(t,t_i)&= \int_{t_i;x_+=0}^{t;x_+=0}Dx_+
\int_{t_i;x_-=0}^{t;x_-=0}Dx_-
\exp\left[\frac{i}{\hbar}{\cal A}[x_+(s),x_-(s)]\right. \cr
&-\left.\frac{1}{\hbar}\int_{t_i}^tds \int_{t_i}^tds'
[x_-(s)\Delta_{cl}(s')
\nu(s,s')]\right]. \cr} \eqno(3.8)
$$
We can write  the classical solutions $\Sigma_{cl}$ and $\Delta_{cl}$
in terms of the elementary functions
$$
\Sigma_{cl}(s)=\Sigma_i u_1(s)+\Sigma_f u_2(s)
                                                        \eqno(3.9.a)
$$
$$
\Delta_{cl}(s)=\Delta_i v_1(s)+\Delta_f v_2(s)
                                                        \eqno(3.9.b)
$$
which satisfy the  boundary conditions
$$
\eqalign{
& u_1(s=t_i)=~1~=~u_2(s=t)\cr
& u_1(s=t)=~0~=~u_2(s=t_i)\cr}\eqno(3.10.a)
$$
$$
\eqalign{
& v_1(s=t_i)=~1~=~v_2(s=t)\cr
& v_1(s=t)=~0~=~v_2(s=t_i).\cr}\eqno(3.10.b)
$$
Now setting
$$
\eqalign{
b_1(t,t_i) &=M(t)\dot{u}_2(t)+\frac{M(t)B(t)}{2},\;\;\;
b_2(t,t_i)=M(t_i)\dot{u}_2(t_i) \cr
b_3(t,t_i) &=M(t)\dot{u}_1(t),\;\;\;b_4(t,t_i)=M(t_i)\dot{u}_1(t_i)
+\frac{M(t_i)B(t_i)}{2}
 \cr}
\eqno(3.11)
$$
where the dot denotes the derivative with respect to $s$ and
$$
a_{ij}(t,t_i)
=\frac{1}{1+\delta_{ij}}\int\limits_{t_i}^tds\int\limits_{t_i}^{t}ds'
 v_i(s)\nu(s,s')v_j(s')
                                                        \eqno(3.12)
$$
we get
$$
\eqalign{
{\cal J}_r(x_f,x'_f,t~|~x_i,x'_i,t_i)& = Z(t,t_i)\exp\left[\frac{i}{\hbar}
 \{b_1\S_f\D_f-b_2\S_f\D_i+b_3\S_i\D_f-b_4\S_i\D_i\}\right] \cr
&\times\exp\left[-\frac{1}{\hbar}\{a_{11}\Delta_i^2
+a_{12}\Delta_i\Delta_f
+a_{22}\Delta_f^2\}\right]. \cr }
                               \eqno(3.13)
$$
The evolution operator (3.13) must preserve the normalisation of the density
matrix. By requiring that $Tr(\rho)=1$, (2.5) implies
$$
\int_{-\infty}^{\infty} dx {\cal J}_r(x,x,t|x_i,x_i',t_i)=\delta(x_i-x_i').
$$
We therefore find that
$$
Z(t,t_i)=\frac{b_2(t,t_i)}{2\pi\hbar}.
\eqno(3.14)
$$

\subsection {Master Equation}

We now proceed with the derivation of the master equation from
the evolution operator (3.13)
using the simplified method of Paz \cite{PazSpain}.
We first take the time derivative of both sides
of (3.13), multiply both sides by $\rho_r(\S_i,\D_i,t_i)$ and
integrate over $\S_i,\D_i$ to obtain
$$
\eqalign{\dot{\rho}_r(\S_f,\D_f,t)&=\left[\frac{\dot{Z}}{Z}+
\frac{i}{\hbar}\dot{b}_1\S_f\D_f-\dot{a}_{22}\frac{\D_f^2}{\hbar}\right]
\rho_r(\S_f,\D_f,t) \cr
&+\frac{i}{\hbar}\D_f\dot{b}_3\int d\D_id\S_i\S_i{\cal
J}_r\rho_r(\S_i,\D_i,t_i) \cr
&-\frac{1}{\hbar}(i\dot{b}_2\S_f+\dot{a}_{12}\D_f)
\int d\D_id\S_i\D_i{\cal J}_r\rho_r(\S_i,\D_i,t_i) \cr
&-\frac{i}{\hbar}\dot{b}_4\int d\D_id\S_i\S_i\D_i{\cal
J}_r\rho_r(\S_i,\D_i,t_i) \cr
&-\frac{\dot{a}_{11}}{\hbar}\int d\D_id\S_i\D_i^2{\cal
J}_r\rho_r(\S_i,\D_i,t_i). \cr}
\eqno(3.15)
$$
Here the dot denotes derivative with respect to $t$.
We can perform the integrals in (3.15) by using
$$
\D_i{\cal J}_r=\frac{i\hbar}{b_2}\frac{\partial {\cal J}_r}{\partial\S_f}
+\frac{b_1\D_f}{b_2}{\cal J}_r
\eqno(3.16a)
$$
$$
\S_i{\cal J}_r=-\frac{i}{b_3}\left[\hbar\frac{\partial {\cal
J}_r}{\partial\D_f}+
(\D_ia_{12}+2\D_fa_{22}){\cal J}_r)\right]-\frac{b_1}{b_3}\S_f{\cal J}_r
\eqno(3.16b)
$$
$$
\eqalign{
\S_i\D_i{\cal J}_r&=-\left(\frac{i\hbar}{b_2}\frac{\partial}{\partial\S_f}
+\frac{b_1\D_f}{b_2}\right) \cr
&\times\left(\frac{i\hbar}{b_3}\frac{\partial}{\partial\D_f}+\frac{i}{b_3}
[\D_ia_{12}+2\D_fa_{22}]+\frac{b_1}{b_3}\S_f\right){\cal J}_r. \cr}
\eqno(3.16c)
$$
The derivation of the master equation simplifies greatly with the use of
the following relations
$$
u_1(s)=w_1(s)-w_2(s)\frac{w_1(t)}{w_2(t)},\;\;
u_2(s)=\frac{w_2(s)}{w_2(t)}.
\eqno(3.17)
$$
In order to satisfy the boundary conditions, (3.10a), we require
$$
w_1(t_i)=\dot w_2(t_i)=1,\;\;\;w_2(0)=\dot w_1(0)=0.
\eqno(3.18)
$$
In this representation we can show that
$$
\frac{\dot{b}_4}{b_2b_3}=-\frac{1}{M(t)},
\;\;\;b_1=-M(t)\frac{\dot{b}_2}{b_2}+M(t)\frac{B(t)}{2},\;\;\;
\dot{a}_{11}=-\dot{v}_1 (t)a_{12}.
\eqno(3.19)
$$
With these relations the master equation reduces to
$$
\eqalign{ i\hbar{\partial\over\partial t}~\rho_r(x,x',t)
=&\biggl\{
  -{\hbar^2\over {2M(t)}}\Bigl({\partial^2\over\partial x^2}-
  {\partial^2\over\partial x'^2}\Bigr) +\frac{i\hbar}{2}B(t)
\Bigl(x\frac{\partial}{x}+x'\frac{\partial}{\partial x'}\Bigr)  \cr
 & +\frac{M(t)}{2}\Omega^2_{ren}(t,t_i)\bigl(x^2 -x'^2 \bigr) +i\hbar
\frac{B(t)}{2}\biggr\}~
\rho_r(x,x',t)~\cr
& -i\hbar\Gamma(t,t_i)(x-x')\Bigl({\partial\over\partial x}
  -{\partial\over\partial x'}\Bigr)~ \rho_r(x,x',t)~ \cr
& +iD_{pp}(t,t_i)(x-x')^2~ \rho_r(x,x',t)~ \cr
& -
\hbar\Bigl(D_{xp}(t,t_i)+D_{px}(t,t_i)\Bigr)(x-x')\Bigl({\partial\over\partial
x}
  +{\partial\over\partial x'}\Bigr) ~\rho_r(x,x',t) \cr
& -i\hbar^2 D_{xx}(t,t_i)\frac{\partial^2}{(\partial x+\partial x')^2}
\rho_r(x,x',t)
\cr}
\eqno(3.20)
$$
where
$$
\Omega_{ren}^2(t,t_i)=\frac{b_1\dot{b}_3}{M(t)b_3}-\frac{\dot{b}_1}{M(t)}
+\frac{B^2(t)}{4}-\frac{\dot{b}_2B(t)}{2b_2}
\eqno(3.21)
$$
$$
\Gamma(t,t_i)=-\frac{1}{2}\left(\frac{\dot{b}_3}{b_3}-
\frac{\dot{b}_2}{b_2}\right)
\eqno(3.22)
$$
$$
D_{pp}(t,t_i)=\frac{b_1^2}{b_2}\left(\frac{a_{12}}{M(t)}-\frac{\dot{a}_{11}}{b_2}\right)+
\frac{2b_1}{M(t)}a_{22}-\dot{a}_{22}+2\frac{\dot{b}_3}{b_3}a_{22}+a_{12}
\frac{b_1\dot{b}_3}{b_2b_3}-\dot{a}_{12}\frac{b_1}{b_2}
\eqno(3.23)
$$
$$
D_{xp}(t,t_i)=D_{px}(t,t_i)=
-\frac{1}{2}\left[\frac{\dot{a}_{12}}{b_2}-2\frac{a_{22}}{M(t)}-
\frac{\dot{b}_3a_{12}}{b_3b_2}
-\frac{2b_1}{b_2}\left(\frac{a_{12}}{M(t)}-\frac{\dot{a}_{11}}{b_2}\right)\right]
\eqno(3.24)
$$
$$
D_{xx}(t,t_i)=\frac{1}{b_2}\left(\frac{a_{12}}{M(t)}-\frac{\dot{a}_{11}}{b_2}\right).
\eqno(3.25)
$$
The dot in these equations denotes the derivative with respect to
$t$.
We can rewrite the master equation in the operator form which may be easier for
physical interpretation. We find that it becomes
$$
\eqalign{
i\hbar\frac{\partial}{\partial t}\hat{\rho}_r (t)= & [\hat{H}_{ren},
\hat{\rho}]+iD_{pp}[\hat{x},[\hat{x},\hat{\rho}]]+iD_{xx}[\hat{p},[\hat{p},
\hat{\rho}]] \cr
& +iD_{xp}[\hat{x},[\hat{p},\hat{\rho}]]+iD_{px}[\hat{p},[\hat{x},\hat{\rho}]]
+\Gamma [\hat{x},\{\hat{p},\hat{\rho}\}] \cr}
\eqno(3.26)
$$
where
$$
\hat{H}_{ren}=\frac{\hat{p}^2}{2M(t)}-
\frac{B(t)}{4}(\hat{p}\hat{x}+\hat{x}\hat{p})+\frac{M(t)}{2}\O_{ren}(t)\hat{x}^2.
\eqno(3.27)
$$
{}From the master equation we know that $D_{xx}$ and $D_{pp}$ generate
decoherence in $p$ and $x$ respectively and  $\Gamma$ gives dissipation.
The master equation differs from Paper I
by more than changing the kernels. The factor
$a_{12}/M(t)-\dot{a}_{11}/b_2$ vanishes only when the dissipation kernel is
stationary (i.e a function of $s-s'$)
and also when the system is a time independent harmonic oscillator. When this
happens
$v_1(s)=u_2(t-s)$ and we have $\dot{v}_1 (t)=-b_2/M(t)$. We see from (3.19)
that the factor
$a_{12}/M(t)-\dot{a}_{11}/b_2$ is zero in this case. All the diffusion
coefficients
contain this factor and $D_{xx}$ depends solely on it. Thus $D_{xx}$ arises
purely from non-stationarity in the dissipation kernel and system.

The coefficients $D_{xx},D_{pp}, D_{xp}$ and $D_{px}$ promote diffusion in the
variables $p^2, x^2$ and $ xp+px$ respectively. This can be seen by going
from the master equation to the Fokker-Planck equation for the Wigner
function \cite{HPZ1,Dekker}. The Wigner function is defined by
$$
F_W(\S,p,t)=\frac{1}{2\pi\hbar}\int_{-\infty}^{\infty}e^{ip\D/\hbar}\langle
\S-{\D \over 2}|\hat{\rho}|\S+{\D \over 2}\rangle d\D.
\eqno(3.28)
$$
where $\S,\D$ are defined in (2.13).
We can show that the Wigner distribution function
from the master equation (3.26-7) (with $B(t)=0$) obeys the following
Fokker-Planck
type equation \cite{Dekker}
$$
\eqalign{
{\partial\over\partial t}F_W(\S,p,t)=&\biggl[
-{p\over M(t)}{\partial\over\partial \S}
+M(t){\Omega_{ren}^2(t)\over 2}~\S{\partial\over\partial p}
+\Gamma(t){\partial\over\partial p}p
-2D_{pp}(t){\partial^2\over\partial p^2} \cr
&\qquad-\hbar D_{xx}(t){\partial^2\over\partial \S^2}
+2\Bigl(D_{xp}(t)+D_{px}(t)\Bigr){\partial^2\over\partial \S\partial p}
\biggr] F_W(\S,p,t).\cr}
                         \eqno(3.29)
$$

\section{Simple Examples}

\subsection{Squeezed Thermal Bath of Static  Harmonic Oscillators}
This is the simplest case treated before in Paper I and II.
In this case the bath modes have time independent coupling constants with
the Lagrangian
\be
L(t)=\frac{1}{2}[\dot{q}^2-\o^2 q^2].
\te
{}From (2.1) $m_n=1, b_n=0$ and $\omega_n^2=\o^2$. Substituting these
into  (2.21) and solving (2.20)  (with $\kappa=\o$) one obtains
\be
\a=e^{-i\o t},\;\;\b=0
\te
where $\a=1$ at the initial time $t=0$. Substituting these into
(2.18-19) one obtains
\be
\mu(s,s')=-\int_0^{\infty}d\o I(\o)\sin \o(s-s')
\te
and
\bea
\nu(s,s') &=&\int_0^{\infty}d\omega\coth\left(\frac{\hbar
\o}{2k_BT}\right)I(\o)
\Bigl[\cosh 2r(\o)\cos[\o(s-s')] \nonumber \\
&-&\sinh 2r(\o)\cos[2\phi(\o)-\o(s+s')\Bigr].
\tea
This is a simple generalisation of previous studies in that we have a squeezed
thermal initial state \cite{sqth} rather than a thermal state.
There are two distinct contributions to the noise kernel for an initially
squeezed bath. The first term represents a change in the spectrum
of the stationary vacuum noise. The second term has a new feature which is
a non-stationary contribution
to the noise kernel. This is expected since the fluctuations of a squeezed
vacuum mode oscillate between conjugate observables.

As $(s+s')\rightarrow\infty$ the second term in (4.4) tends to zero. Thus
the nonstationarity in the noise kernel is a transient effect for the initial
squeezed bath. For an initial squeezed bath with thermal spectrum
the late time noise kernel would tend to that of the usual thermal state.
This is because  at late times , the noise kernel $\nu$
loses track  of the initial phase distribution $\theta(\omega)$.
This is, however,  not true for the master equation
diffusion coefficients. Equations (3.23-25) show that the diffusion
coefficients depend
on the noise kernel in a non-local way in time. It may be interesting to
compare the timescales in which the semi-classical system and the full
quantum system forget the $\phi(\omega)$ initial condition in the bath.

Although we have considered only single mode squeezed initial
states our results can be easily extended to two-mode squeezed initial
states \cite{sqst}. This will change the noise kernel (4.4) but not the
dissipation
kernel (4.3) which remains independent of the initial state.
Since the influence functional
(2.10) is unchanged the exact forms for the evolution operator and
master equations in Sec. 3 will stay. These results could then be
used for an accurate description of systems
coupled to an initially squeezed bath\cite{milburn,sqth}.

If we set the initial squeezing to zero we regain the results of Paper I.
For completeness we will summarise the simple analytical results obtained
previously. In this case
the noise and dissipation kernels
are  functions only of $s-s'$. They can always be related by some integral
equation known as the fluctuation--dissipation relation (FDR) \cite{HPZ1}:
\be
\nu(s)=\int\limits_{-\infty}^{+\infty}ds'~K(s-s')~\gamma(s')
\te
where the kernel $K(s)$ is
\be
K(s)=\int\limits_0^{+\infty}{d\omega\over\pi}~
     \omega\coth{1\over 2}\beta\hbar\omega~\cos\omega s
\te
and $\mu(s)=\frac{d}{ds}\gamma(s)$.
In the classical or high temperature limit, the kernel $K$ is
proportional to the delta function $K(s)=2k_BT\delta(s)$ and the FDR is
equivalent to the well known Einstein formula.

An interesting case is
an environment which generates an ohmic spectral density
\be
I(\o)=\frac{2}{\pi}\g_0 M\o.
\te
With a discrete high frequency cutoff $\Lambda$,
\bea
\mu(s)&= &\frac{2}{\pi}\g_0 M\frac{d}{ds}\left(\frac{\sin \Lambda s}{s}\right)
\nonumber \\
&\rightarrow &2\g_0 M\frac{d}{ds}\delta(s),\;\; as ~ \Lambda\rightarrow\infty.
\tea
In this case for a harmonic oscillator system (2.17) becomes
\be
\ddot{X}(s)+2\g_0\dot{X}+\Omega_r^2 X=-\bar{\xi}(t)
\te
where $\Omega_r=\Omega-\frac{4}{\pi}\g_0\Lambda$.
We see that the ohmic environment is special in that it gives local dissipation
in the infinite cutoff limit.

Theoretically, the meaning of renormalization can be understood
as follows \cite{HPZ1}: We can rewrite the action as
\be
S = \int^{t}_{0} ds
\Bigl\{ {1\over 2} M(\dot{x}^2 - \Omega^2 x^2) +
\sum_{n} [{1\over 2} m_n \dot q_n^2
- {1\over 2} m_n\omega_n^2
(q + {c_n \over m_n\omega_n^2} x )^2
+ {1\over 2} {c^2_n \over m_n\omega_n^2}
x^2].\Bigr\}
\te
The last term can be viewed as a frequency counter term $\Omega_c^2$
arising from the interaction of the Brownian particle with the bath
oscillators
\be
\Omega^2_c
= -{1\over 2M} \sum_{n} {c_n^2 \over m_n\omega_n^2}
= - \int d\omega {I(\omega)\over\omega}.
\te
The bare frequency $\Omega^2$ is thus modified into a renormalized
frequency $\Omega_r^2$ given by
\be
\Omega_r^2 = \Omega^2 + \Omega_c^2.
\te

Another interesting case is the high temperature limit.
 If we consider the temperature to be such that
$\frac{\hbar}{k_B T}<<\Lambda^{-1}$ and then
let $\Lambda\rightarrow\infty$ (the order in which
we take the limits is important), the noise
kernel (4.4) is simplified to
\be
\nu(s)
 ={{4Mk_BT\gamma_0}\over{\hbar}}~\delta(s).
\te
In this case we see that the noise is white with an amplitude
$4\gamma_0Mk_{B}T$,
and (4.9) reduces to the Nyquist formula. In the ohmic, high temperature
and infinite cutoff limit the master equation coefficients can be calculated.
Using (3.5) we find that, for a time independent harmonic oscillator system,
$u_1$ and $u_2$ must satisfy
\be
\ddot{u}(s)+2\g_0\dot{u}(s)+\Omega_r^2 u(s)=-4\g_0\delta(s)u(0).
\te
The solutions satisfying the appropriate boundary conditions (with $t_i=0$)
are
\be
u_1(s)=-\frac{\sin [\tilde{\Omega}(s-t)]e^{-\gamma_0 s}}{\sin
\tilde{\Omega}t},\;\;
u_2(s)=\frac{\sin [\tilde{\Omega}s] e^{-\gamma_0 (s-t)}}{\sin \tilde{\Omega}t}
\te
where $\tilde{\Omega}^2=\Omega_r^2-\gamma_0^2$.
Applying these to (3.11) we find
\bea
b_2(t)&=&\frac{M\tilde{\Omega}e^{\gamma_0 t}}{\sin \tilde{\Omega}t},\;\;
b_3(t)=-\frac{M\tilde{\Omega}e^{-\gamma_0 t}}{\sin \tilde{\Omega}t} \\
b_4(t)&=&-b_1(t)=M(\gamma_0-\tilde{\Omega}\cot \tilde{\Omega}t).
\tea
Since $b_4$ is discontinuous before and after $t=0$ (due to the kick) we have
taken the average.

The results (4.16-17) are exact in the infinite cutoff limit of an ohmic
environment. This is a local approximation which has been shown to be
good for timescales greater than the inverse cutoff \cite{envdec}.
Equations (4.16-17) depend only on the dissipation
kernel which is unchanged by initial squeezing in the bath. Thus these
equations can also be applied to more general situations.

Using the noise kernel (4.13) and the fact that
$v_1(s)=u_2(t-s),\;\;v_2(s)=u_1(t-s)$ we can calculate $a_{ij}$ and
find that the master equation coefficients to be
\be
\Omega_{ren}(t)=\Omega_r,\;\; \Gamma(t)=\gamma_0,\;\;D_{xp}(t)=D_{xx}(t)=0,\;\;
D_{pp}(t)=-\frac{2\g_0k_{B}TM}{\hbar}.
\te
For decoherence studies under these and other environmental conditions see
\cite{envdec}.

\subsection{Bath of Upside Down Oscillators}
This is the next simplest case.
In this case the bath modes have the Lagrangian.
\be
L(t)=\frac{1}{2}[\dot{q}^2+\o^2q^2].
\te
{}From (2.1) $m_n=1, b_n=0$ and $\omega_n^2=-\o^2$. Substituting these
into (2.21) and solving (2.20) (with $\kappa=\o$) we obtain
\be
\a_{\o}(t)=\cosh \o t,\;\;\b_{\o}(t)=-i\sinh \o t
\te
where $\a=0$ at $t=0$ which is our initial time. Substituting these into
(2.18-19) we obtain
\be
\mu(s,s')=-\int_0^{\infty}d\o I(\o)\sinh \o(s-s')
\te
and
\bea
\nu(s,s')&=&\int_0^{\infty}d\o \coth\left(\frac{\hbar \o}{2k_BT}\right)I(\o)
\Bigl[\cosh 2r(\o)\cosh \o(s+s') \nonumber \\
&-&\sinh 2r(\o)\cos 2\phi(\o)\cosh \o(s-s') \nonumber \\
&-&\sinh 2r(\o)\sin 2\phi(\o)\sinh \o(s+s')\Bigr].
\tea
This case can be used as an amplifier model in quantum optics and
electronics \cite{gilson}.

\section{Particle Detector in a Scalar Field Bath}

The formalism developed here can be used to study quantum statistical
processes in cosmological and black hole spacetimes. The
model (2.1) can be used to depict a particle detector in motion,
or an observer near a black hole. It can also be used  to describe
the non-equilibrium dynamics of homogeneous and inhomogeneous modes
(density fluctuations) of the inflaton field
or gravity wave perturbations (which in the linear approximation
obey the  wave equation of a massless, minimally coupled scalar field)
in the early universe.

In this section we will show how a general real scalar field in an expanding
universe is reduced to a sum over quadratic time dependent Hamiltonians.
The action for a free massive ($m$) scalar field in a curved spacetime
with metric $g_{\m\n}$ and scalar curvature $R$ is given by
\begin{equation}
S=\int {\cal L}(x)d^4x= \int\frac{\sqrt{-g}}{2}d^4x
\left(g^{\mu\nu}\bigtriangledown_{\mu}\Phi\bigtriangledown_{\nu}\Phi-
(m^2+\xi_d  R)\Phi^2\right)
\end{equation}
where $\bigtriangledown_\nu$ denotes covariant derivative,
and $\xi_d$ is the field coupling
constant ($\xi_d=0, 1/6$ respectively for minimal and conformal coupling).
In the spatially-flat Robertson-Walker (RW) metric
\begin{equation}
ds^2=a^2(\eta)[d\eta^2-\sum_i dx^2_i]
\end{equation}
we can write
\begin{equation}
{\cal L}(x)=\frac{1}{2}a^2(\eta)\left[(\dot{\Phi})^2-\sum_{i}(\Phi_{,i})^2-
\left(m^2 a^2+  6\xi_d  \frac{\ddot{a}}{a}\right)\Phi^2\right]
\end{equation}
where a dot denotes derivative taken with respect to conformal
time $\eta=\int dt/a$.
If we rescale the field variable $\chi=a\Phi$, this becomes
\begin{equation}
{\cal L}(x)=\frac{1}{2}\left[(\dot{\chi})^2-\sum_{i}(\chi_{,i})^2-
\left(m^2 a^2+(6\xi_d -1)\frac{\ddot{a}}{a}\right)\chi^2-
(1-6\xi_d )\frac{d}{d\eta}\left(\frac{\dot{a}}{a}\chi^2\right)\right]
\end{equation}
where the last term is a surface term. \footnote{The part of the surface term
proportional to $\xi_d$ has been added in by hand. The surface term
ensures that the second derivative of the scale factor doesn't appear in the
Lagrangian \cite{Mat}.
This is necessary to have a consistent variational theory when
the scale factor is treated dynamically rather than kinematically \cite{Sim}.}

If we confine the scalar field in a box of co-moving volume $L^3$
(fixed coordinate volume), we can expand it in normal modes
\begin{equation}
\chi(x)=\sqrt{\frac{2}{L^3}}\sum_{\vec{k}}[q_{\vec{k}}^+
 \cos\vec{k}\cdot\vec{x} + q_{\vec{k}}^- \sin\vec{k}\cdot\vec{x}]
\end{equation}
which leads to the Lagrangian
\begin{equation}
L (\eta)=\frac{1}{2}\sum_{\sigma}^{+-}\sum_{\vec{k}}
\left[(\dot{q}_{\vec{k}}^{\sigma})^2-2(1-6\xi_d )\frac{\dot{a}}{a}
q_{\vec{k}}^{\sigma}\dot{q}_{\vec{k}}^{\sigma}
-\left(k^2+m^2 a^2+(6\xi_d -1)\frac{\dot{a}^2}{a^2}
\right)q_{\vec{k}}^{\sigma2}\right]
\end{equation}
where $k=|\vec{k}|$ and $L(\eta)=\int{\cal L}(x)d^3 \vec{x}$.
Canonical momenta are
\begin{equation}
p_{\vec{k}}^{\sigma}=\frac{\partial L (\eta)}{\partial\dot{q}_{\vec{k}}^
{\sigma}}=\dot{q}_{\vec{k}}^{\sigma}-(1-6\xi_d)\frac{\dot{a}}{a}q_{\vec{k}}^{\sigma}.
\end{equation}
Defining the canonical Hamiltonian the usual way we find
\begin{equation}
H (\eta)=\frac{1}{2}\sum_{\sigma}^{+-}\sum_{\vec{k}>0}
\left[p_{\vec{k}}^{\sigma 2}+(1-6\xi_d )\frac{\dot{a}}{a}(p_{\vec{k}}
^{\sigma}q_{\vec{k}}^{\sigma}+q_{\vec{k}}^{\sigma}p_{\vec{k}}^{\sigma})
+\left(k^2+m^2 a^2+6\xi_d (6\xi_d -1)\frac{\dot{a}^2}{a^2}\right)q_{\vec{k}}
^{\sigma 2}\right]
\end{equation}
where the sum is over positive $\vec k$ only since we have an expansion
over standing rather than travelling waves.

The system is quantized by promoting $(p_{\vec{k}}^{\sigma},q_{\vec{k}}
^{\sigma}),\;(p_{s\vec{k}}^{\sigma},q_{\vec{k}}
^{\sigma})$
to operators obeying the usual harmonic oscillator commutation relation.
Thus the amplitude functions of the normal modes behave like time-dependent
harmonic oscillators. (The Hamiltonian is not unique but is a result of our
time coordinate and choice of canonical variables.)

The above shows that a scalar field
can be represented as a bath of parametric oscillators.  In order to
study the noise properties of the quantum field, we now introduce an
interaction between the system, which can be a particle detector
or a field mode, and the bath, the scalar field.

\subsection{Spectral Density of a Scalar Field}

Consider the general form of
interaction between the system harmonic oscillator $r$,
and a scalar field $\chi$ of the form
\be
{\cal L}_{int}(x)=-\e r\chi(x)\delta(\vec{x}_0).
\te
They are coupled at the spatial point $\vec{x}_0$
with coupling strength  $\e$. We want to derive the spectral density function
for this field $ I(\o) = \sum \d (\o-\o_n) c_n^2/2\kappa_n$.
Integrating out the spatial variables we find that
\be
L_{int}(\eta)=\int{\cal L}_{int}(x)d^3\vec{x}=-\e r\chi(\vec{x}_0,\eta)
\te
where
\be
\chi(\vec{x}_0,\eta)=\sqrt{\frac{2}{L^3}}\sum_{\vec{k}}[q_{\vec{k}}^+
 \cos\vec{k}\cdot\vec{x}_0 + q_{\vec{k}}^- \sin\vec{k}\cdot\vec{x}_0].
\te
Comparing this with (2.1) we see that each set of modes has the effective
coupling constants
\be
c_{\vec k}^+ = \sqrt{\frac{2}{L^3}}\e\cos\vec{k}\cdot\vec{x}_0, ~~~
c_{\vec k}^- = \sqrt{\frac{2}{L^3}}\e\sin\vec{k}\cdot\vec{x}_0.
\te
In the continuous limit the oscillator label $n$ is replaced by $\vec{k}$.
Adding the spectral densities from both the $\pm$ sets of modes we obtain
\be
I(k)=\frac{\e^2}{L^3}\sum_{\vec{k}}\delta(k)
\frac{1}{\kappa_k}
\te
where $\omega$ is replaced by $k$.
In the continuous limit:
$\sum_{\vec{k}}\rightarrow\left(\frac{L}{2\pi}\right)^3\int d^3\vec{k}$.
Writing $d^3\vec{k}=k^2\sin \theta dkd\theta d\phi$
and integrating between the limits $\phi[2\pi,0]$ and $\theta[\pi/2,0]$
(remembering we only include half of the modes)
$\sum_{\vec{k}} \rightarrow
\frac{L^3}{(2\pi)^2}\int_0^{\infty}k^2 dk$, we get
\be
I(k)=\frac{\e^2}{(2\pi)^2}\frac{k^2}{\kappa_k}.
\te
For a two-dimensional scalar field we get
$$
I(k)=\frac{\e^2}{2\pi \kappa_k}.
$$

\subsection{Accelerated Observer}
We consider a two dimensional massless scalar field $\Phi$ in
flat space with mode decomposition
\begin{equation}
\Phi(x)=\sqrt{\frac{2}{L}}\sum_{k}[q_{k}^+
\cos kx + q_{k}^- \sin kx].
\end{equation}
The Lagrangian for the field can be expressed as a sum of coupled oscillators
with amplitudes $q^{\pm}_k$ for each mode
\be
L (s)=\frac{1}{2}\sum_{\sigma}^{+-}\sum_{k}
\left[(\dot{q}_{k}^{\sigma})^2
-k^2 q_k^{\sigma2}\right].
\end{equation}
Now consider an observer undergoing constant acceleration $a$ in this field
with trajectory
\be
x(\t)=\frac{1}{a}\cosh a\t,\;\;\;s(\t)=\frac{1}{a}\sinh a\t.
\te
We want to show via the influence functional method that the observer
detects a thermal radiation. This effect was first proposed by Unruh
\cite{Unr}, as inspired by the Hawking effect \cite{Haw75} for black holes.
Let us see what the spectral density is. The particle- field interaction is
\be
{\cal L}_{int}(x)=-\e r\Phi(x)\delta(x(\t))
\te
where they are coupled at the spatial point $x(\t)$ with coupling strength
$\e$
and $r$ is the detector's internal coordinate.
Integrating out the spatial variables we find that
\be
L_{int}(\t)=\int{\cal L}_{int}(x)dx=-\e r\Phi (x(\t)).
\te
Comparing (5.19) with (2.1) we see that the accelerated observer is coupled
to the field with  effective coupling constants
\be
c_n^+(s)=\e \sqrt{\frac{2}{L}}\cos kx(\t),\;\;c_n^-(s)=\e
\sqrt{\frac{2}{L}}\sin kx(\t).
\te
{}From (2.26) the spectral density in the discrete case is given by
\be
I(k,\t,\t')=\sum_{\sigma}^{+-}\sum_{n}\frac{\delta(k-k_n)c_n^{\sigma}(\t)
c_n^{\sigma}(\t')}{2\omega_n}
\te
where we have to include the sum over both sets of modes
and we have put $\kappa_n=\omega_n=|k_n|$. This ensures
that $f_n(s_i)=0$ in (2.21). Making use of (5.20) and $\sum_n\rightarrow
\frac{L}{2\pi}\int dk$
we find that (5.21) becomes
\be
I(k,\t,\t')=I(k)\cos k[x(\t)-x(\t')]
\te
where $I(k)=\frac{\e^2}{2\pi\omega}$ is the spectral density of the (2-dim)
scalar field seen by an inertial detector.
{}From (4.3) and (4.4) we can write, using an initial vacuum state,
\be
\zeta(s(\t),s(\t'))=\nu(s,s')+i\mu(s,s')=\int_0^{\infty}
dk~I(k,\t,\t')e^{-i\omega[s(\t)-s(\t')]}.
\te
We can rewrite this as
\bea
\zeta(\t,\t') &=&\frac{1}{2}\int_0^{\infty}dk~I(k)
e^{-ik[x(\t)-x(\t')+s(\t)-s(\t')]} \nonumber \\
&+&\frac{1}{2}\int_0^{\infty}dk~I(k)
e^{-ik[x(\t')-x(\t)+s(\t)-s(\t')]}
\tea
which upon using (5.17) can be written as
\bea
\zeta(\t,\t')&=&\frac{1}{2}\int_0^{\infty}dk'~I(k')
\Bigl[\exp\left(-2ik'e^{a\Sigma}\sinh [a\Delta]/a\right)
\nonumber \\
&+&\exp\left(-2ik'e^{-a\Sigma}\sinh [a\Delta]/a\right)\Bigr]
\tea
where $2\Sigma=\t+\t'$ and $\Delta=\t-\t'$.
Making use of \cite{Ang}
\be
e^{-i\alpha\sinh (x/2)}=\frac{4}{\pi}\int_0^{\infty}d\nu
K_{2i\nu}(\alpha)[\cosh (\pi\nu)\cos (\nu x)-i\sinh (\pi\nu)\sin (\nu x)]
\te
where $K_n$ is a Bessel function of order $n$, we find that (5.25) becomes
\be
\zeta(\t,\t')=\int_0^{\infty}dk~G(k)\left[\coth \left(\frac{\pi k}{a}\right)
\cos k(\t-\t')-i\sin k(\t-\t')\right]
\te
where
\bea
G(k)&=&\frac{2}{\pi a}\sinh (\pi k/a)\int_0^{\infty}dk'~I(k')\Bigl[K_{2ik/a}
\left(2k'
e^{a\Sigma}/a\right) \nonumber \\
&+&K_{2ik/a}\left(2k'
e^{-a\Sigma}/a\right)\Bigr] =I(k).
\tea
In deriving this we have used the integral identity
\be
\int_0^{\infty}dx~x^{\mu}K_{\nu}(ax)=2^{\mu-1}a^{-\mu-1}\Gamma
\left(\frac{1+\mu+\nu}{2}\right)\Gamma\left(\frac{1+\mu-\nu}{2}\right)
\te
and the properties of gamma functions.
Comparing (5.27) with (4.3-4) we see that
a thermal spectrum is detected at temperature
\be
k_B T= {a\over {2 \pi }}.
\te

This was first found by Unruh \cite{Unr}
and stated in this form recently by Anglin \cite{Ang}.

\subsection{Hawking Radiation in Black Holes}
Consider the metric of a two-dimensional uncharged black hole with mass $m$
\be
ds^2=\left(1-\frac{2m}{r}\right)dt^2-\left(1-\frac{2m}{r}\right)^{-1}dr^2.
\te
In the Regge-Wheeler coordinates
\be
r^*=r+2m\ln|r/(2m)-1|
\te
this can be written as
\be
ds^2=\left(1-\frac{2m}{r}\right)(dt^2-dr^{*2}).
\te
The Kruskal coordinates are defined by
\be
\bar{t}-\bar{r}^*=-4m \exp\left[\frac{r^*-t}{4m}\right],\;\;\bar{t}+\bar{r}^*=
4m\exp\left[\frac{r^*+t}{4m}\right].
\te
With this the metric becomes
\be
ds^2=\frac{2m}{r}e^{-r/(2m)}(d\bar{t}^2-d\bar{r}^{*2}).
\te

Since the metric (5.35) is conformal to flat space, the field theory is
equivalent to that of flat space. Thus a detector at constant Kruskal
position $\bar{r}^*$ will have an influence functional identical in
form to that of an inertial detector in flat two-dimensional spacetime
in Kruskal coordinates.
However we are interested in a detector at constant $r^*$. In this case we see
from (5.34) that constant $r^*$ is effectively an accelerating detector in
Kruskal coordinates since
\be
\bar{r}^*(t)=4me^{r^*/(4m)}\cosh [t/(4m)].
\te
We also want to express the influence functional in cosmic time $t$
which from (5.34) is
\be
\bar{t}(t)=4me^{r^*/(4m)}\sinh [t/(4m)]
\te
for the detector at constant $r^*$.
This case is now similar to the accelerating observer
and as in Sec. 5.2 the spectral density is
\be
I(k,t,t')=I(k)\cos k[\bar{r}^*(t)-\bar{r}^*(t')]
\te
where $I(k)=\frac{\e^2}{2\pi\omega}$ and $\omega=|k|$.
With this spectral density we can write for a massless scalar field in a
two-dimensional black hole spacetime
\bea
\zeta(t,t') \equiv
\nu(t,t')+i\mu(t,t')&=&\frac{1}{2}\int_0^{\infty}dk~I(k)
e^{-ik[\bar{r}^*(t)-\bar{r}^*(t')+\bar{t}(t)-\bar{t}(t')]} \nonumber \\
&+&\frac{1}{2}\int_0^{\infty}dk~I(k)
e^{-ik[\bar{r}^*(t')-\bar{r}^*(t)+\bar{t}(t)-\bar{t}(t')]}.
\tea
Comparing (5.39) and (5.24) we see that this case is identical to
the accelerated observer if we identify $a\equiv 1/(4m)$. The factor involving
$r^*$ can be absorbed into the definition of $k$.
Hence we can rewrite (5.39) as
\be
\zeta(t,t')=\int_0^{\infty}dk~I(k)\left[\coth (4\pi mk)
\cos k(t-t')-i\sin k(t-t')\right].
\te
Comparing (5.40) with (4.3-4) we see that
a thermal spectrum is detected by an observer at constant $r^*$ at temperature
\be
k_B T= {1\over {8 \pi m}}.
\te

In the two dimensional case the detector response is independent of its
position $r^*$. This will not be the case in four dimensions.

\subsection{Hawking Radiation in de Sitter Space}

We now illustrate how the Gibbons-Hawking result \cite{GibHaw} can be
obtained from the influence functional method.
These examples are also of practical use for describing the non-equilibrium
dynamics of the homogenous and inhomogenous (density fluctuations) modes
of the inflaton field in the early universe
\cite{inflation,stoinf,galforinf,GutPi,CorBru,decinf,Mat}.

\subsubsection{Massless conformally coupled field}
Consider now a four-dimensional spatially-flat Robertson-Walker (RW)
spacetime with metric
\be
ds^2= dt^2 -\sum_i a^2(t)dx_i^2.
\te
For this metric the Lagrangian density of a massless conformally coupled scalar
field,
defined by (5.1), is
\begin{equation}
{\cal
L}(x)=\frac{a^3}{2}\left[(\dot{\Phi})^2-\frac{1}{a^2}\sum_{i}(\Phi_{,i})^2
-\left(\frac{\dot{a}^2}{a^2}-\frac{\ddot{a}}{a}\right)\Phi^2\right]
\end{equation}
where a dot denotes a derivative with respect to $t$.
Decomposing $\Phi$ in standing wave normal modes we find (after adding a
surface term)
\begin{equation}
L(t)=\int{\cal L}(x)d^3 \vec{x}=\sum_{\sigma}^{+-}\sum_{\vec{k}}\frac{a^3}{2}
\left[(\dot{q}_{\vec{k}}^{\sigma})^2+2\frac{\dot{a}}{a}\dot{q}_{\vec{k}}^{\sigma}q_{\vec{k}}^{\sigma}-
\left(\frac{k^2}{a^2}-\frac{\dot{a}^2}{a^2}\right) q_{\vec{k}}^{\sigma
2}\right]
\end{equation}
where $k$=$|\vec{k}|$.
If we wrote the Lagrangian in terms of conformal rather than cosmic time we see
from (5.6) that we would have obtained a bath of stationary oscillators. Our
kernels would then be (4.3) and (4.4) but written in conformal time.
If we were to rewrite these kernels in cosmic time we would get the same
kernels
as those by starting with a Lagrangian written in cosmic time as we are doing
here.

The detector-field interaction is of the same form as (5.9) (with $\chi$
replacing $\Phi$)
and we find that with $\kappa_k=k$
(5.14) gives
\be
I(k)=\left(\frac{\e}{2\pi}\right)^2 k.
\te
Using the Lagrangian (5.44) we find from (2.20) that the Bogolubov coefficients
are
\be
\alpha=\frac{(1+a^2)}{2a}e^{-ik\eta},\;\;\beta=\frac{(1-a^2)}{2a}e^{-ik\eta}
\te
where $\eta=\int_{t_i}^t dt/a(t)$ with $a(t_i)=1$. Using these we find
that the noise and dissipation kernels (2.18-19) are, for an initial vacuum
state
\be
\zeta(t,t')=\nu(t,t')+i\mu(t,t')=\frac{1}{a(t)a(t')}\int_0^{\infty}dk
{}~I(k)e^{-ik(\eta-\eta')}.
\te
We will now specialise to the de Sitter dynamics where, in the spatially-flat
RW coordinatization \cite{BirDav}, the scale factor has the time-dependence
\be
a(t)=e^{Ht}.
\te
In this case $\eta=-\frac{1}{H}e^{-Ht}$ with $t_i=0$. If we define
$\Delta=t-t',\;2\Sigma=t+t'$ we find that (5.47) becomes
\be
\zeta(t,t')=e^{-2H\Sigma}\int_0^{\infty}dk ~I(k) \exp
\left[-\frac{2ik}{H}e^{-H\Sigma}\sinh (H\Delta/2)\right].
\te
Using (5.26) we find that
\be
\zeta(t,t')=\int_0^{\infty}dk~G(k)\left[\coth \left(\frac{\pi k}{H}\right)
\cos k(t-t')-i\sin k(t-t')\right]
\te
where
\bea
G(k)&=&\frac{4\sinh (\pi k/H)}{\pi H e^{2H\Sigma}}\int_0^{\infty}dk'~
I(k')K_{2ik/H} (2k'e^{-H\Sigma}/H) \nn \\
&=&\left(\frac{\e}{2\pi}\right)^2 k=I(k).
\tea
We have again used the integral identity (5.29) and the properties
of gamma functions.
Comparing (5.50) with (4.3-4) we see that
a thermal spectrum is detected by an inertial observer in de Sitter space at
temperature
\be
k_B T= {H \over {2 \pi}}.
\te

Cornwall and Bruinsma \cite{CorBru} who considered the evolution of low
momentum
modes of an inflaton field coupled to a thermal bath in a de Sitter background
also derived the influence functional for a conformally coupled scalar field
in de Sitter space. The noise and dissipation kernels they found in their
Eq. (3.31)
differs from ours since they did not add a surface term proportional to
$\xi_d$.
As a result they got
nonstationary kernels when written in conformal time. As we pointed out
previously \cite{Mat}
a surface term is needed to give a consistent
variational theory when the scale factor is treated as a dynamical variable.
In this case we see from (5.6) that in conformal time
conformal coupling with a bath of ordinary stationary oscillators  gives
the usual stationary kernels. In cosmic time these kernels lead to (5.50) which
is still stationary,
but shows the expected Gibbons-Hawking temperature.

\subsubsection{Massless minimally coupled field}
{}From (5.6) the Lagrangian for a minimally coupled massless field in de Sitter
space is
\begin{equation}
L (\eta)=\sum_{\sigma}^{+-}\sum_{\vec{k}}\frac{1}{2}
\left[(\dot{q}_{\vec{k}}^{\sigma})^2+\frac{2}{\eta}\dot{q}_{\vec{k}}^{\sigma}
q_{\vec{k}}^{\sigma}-
\left(k^2-\frac{1}{\eta^2}\right) q_{\vec{k}}^{\sigma 2}\right].
\end{equation}
Solving (2.20) (with $\kappa_n=k$) we find that the Bogolubov coefficients are
\be
\a(\eta)=\left(1-\frac{i}{2k\eta}\right)e^{-ik\eta},\;\;\;
\b(\eta)=-\frac{i}{2k\eta}e^{-ik\eta}.
\te
Substituting these into (2.18-19) we find that
\be
\zeta(\eta,\eta')=\nu(\eta,\eta')+i\mu(\eta,\eta')=\int_0^{\infty}dk ~I(k)
e^{-ik(\eta-\eta')}\left(\frac{1+k^2\eta\eta'+ik(\eta-\eta')}{k^2\eta\eta'}\right)
\te
where $I(k)$ is given by (5.45). We want to write this in terms of cosmic
time given by $\eta=-\frac{1}{H}e^{-Ht}$. Following a similar procedure as
before, we find
\be
\zeta(t,t')=\int_0^{\infty}dk~G(k)\left[\coth \left(\frac{\pi k}{H}\right)
\cos k(t-t')-i\sin k(t-t')\right]
\te
where
\be
G(k)=I(k)\left[1+\frac{H^2}{k^2}+2i\frac{H}{k}\sinh
\left(\frac{H(t-t')}{2}\right)
\tanh \left(\frac{\pi k}{H}\right)\right]
\te
and we have ignored a factor $e^{H(t+t')}$ which gets cancelled by
changing the integration measure from $\eta$ to $t$ in the influence
functional.

The imaginary part of (5.57) generates a contribution to the dissipation
kernel of the form
\be
\mu_{im}(t-t')=\frac{\e^2H}{2\pi}\sinh
\left(\frac{H(t-t')}{2}\right)\delta(t-t').
\te
Inserting this into the influence functional (2.10) we see that it leads to a
vanishing contribution to the influence functional. Similarly the imaginary
part of (5.57) generates a contribution to the noise kernel of the form
\bea
\nu_{im}(t-t')&=&\frac{2H\e^2}{(2\pi)^2}\sinh \left(\frac{H(t-t')}{2}\right)
\int_{0}^{\infty}dk~\tanh\left(\frac{\pi k}{H}\right)\sin k(t-t')\nonumber \\
&=&\frac{2H\e^2}{(2\pi)^2}\left[-\sinh \left(\frac{H(t-t')}{2}\right)
\frac{\cos \Lambda (t-t')}{t-t'}\Bigl|_{\Lambda\rightarrow\infty}~
+\frac{H}{2}\right]
\tea
where we have first integrated by parts and then used a standard integral.
The first term in (5.59) will generate a vanishing contribution to the
influence
functional (2.10) since it involves an integral over a term oscillating
infinitly fast.
The second term in (5.59) can also be ignored since it generates only
a zero frequency contribution to the noise spectrum. Thus the imaginary part of
(5.57) can be
ignored leaving a thermal influence functional at the Gibbons-Hawking
temperature but with an effective spectral density of the form
\be
G(k)=I(k)\left[1+\frac{H^2}{k^2}\right].
\te
We see in this spectral density the well known infrared divergence associated
with massless, minimally coupled fields in de Sitter spacetime.

Habib and Kandrup claimed  \cite{HabKan}, from a classical analysis, that
a fluctuation-dissipation relation (FDR) would increasingly fail to hold
as the physical period
of oscillation increased over the expansion timescale of the universe.
We suspect that the definition of FDR and its applicability in their work
is more restricted than ours.
We see that in both of these examples here the FDR (4.5) is exact
despite the fact that
the physical period of oscillation can be arbitrarily greater than the
expansion timescale. This is consistent with the view of
\cite{HuPhysica,HPZ1,HPZ2} that the FDR is a categorical relation as it
is a description of the full backreaction of the environment on the system.

\section{Discussion}

Many physical problems can be modeled by a quantum Brownian particle
in a parametric oscillator bath. We mention quantum optics, quantum and
semiclassical cosmology
and gravity. This paper aims to accomplish two goals:\\

\noindent I. To derive the influence functional of a parametric oscillator
bath,
the evolution operator and the master equation for the reduced density matrix
for
explicit use in these problems. \\
II. To relate the quantum mechanics of oscillators to quantum fields,
thus providing a bridge from quantum stochastical mechanics to
quantum field theory. This connection can benefit the former with
the well-developed techniques of field theory (e.g.,
use of Feynman diagrams \cite{HPZ2,SinHu,cgea}) and enrich the latter with
imparting a statistical mechanics meanings to many quantum effects
 \cite{HuWaseda,HuTsukuba,HuPhysica}.\\

Two issues are discussed in this paper:\\
A. The nature and origin of noise and dissipation in quantum fields\\
B. The statistical mechanical meaning of quantum processes in the early
universe
and black holes.

On the first issue we have discussed these problems:\\
1) {\it How to extract the statistical information of a quantum field}.
We couple a particle detector to the oscillator bath
and study the detector's response to the fluctuations of the field.
We found that the characteristics of quantum noise vary
with the nature of the field, the type of coupling between the field and the
background spacetime, and the time-dependence of the scale factor of the
universe.\\
2) {\it How to relate noise to particle creation}.
Parametric amplification of vacuum fluctuations and backscattering of waves
in the second-quantized formulation give rise to particle creation.
By writing the influence functional in terms of the Bogolubov
coefficients which determine the amount of particles produced,
one can identify the origin of
noise in this system to particle creation \cite{HMLA,CalHuSG,HM3}.

On the second issue, we have studied the problem of\\
3) {\it quantum noise and thermal radiance}.
We illustrate how a uniformly accelerating detector in Minkowski space,
a static detector outside a black hole
and a comoving observer in a de Sitter universe
observes a thermal spectrum. The viewpoint of quantum open systems
and the method of influence functionals can, in our opinion,
lead to a deeper understanding of black hole thermodynamics and
quantum processes in the early universe \cite{HuWaseda}.\\

As further studies, the results obtained here can be useful for the
following problems:\\
a) {\it Decoherence}. The transition of the system from quantum to classical
requires the diminishing of coherence in the wave function.
The noise kernel is found to be primarily responsible for this decoherence
process.
Decoherence can be studied by analyzing the magnitude of the diffusion
coefficients
in the master equation. The new result obtained here
is useful for the analysis of decoherence where parametric excitation
is present in the environment. This is the case when considering the quantum
to classical transition of the wavefunction of the universe
\cite{decQC,PazSin},
homogenous and inhomogenous modes (density fluctuations)
of an inflaton field \cite{decinf,Mat,HuBelgium} or
the primordial gravitational radiation background.
For the case of density fluctuations we can expect decoherence, dissipation
and diffusion to have important consequences for the amplitude and spectrum of
density perturbations.
The relation of particle creation and decoherence was one of the original
physical motivations for this work. Indeed one of us has speculated
\cite{HuTsukuba} that in the early universe, vacuum particle creation
and decoherence can be important at the same scale near the Planck time.
We will address these issues at a later time.\\
b) {\it Backreaction} The backreaction of these quantum field processes
manifests as dissipation effect, which is described by the dissipation
kernel  \cite{dissip}.
In \cite{HM3,CalHuSG} we outline a program for studying the backreaction of
particle creation in semiclassical cosmology in the open system framework.
We use a model where the quantum Brownian particle
and the oscillator bath are coupled parametrically. The field parameters
change in time through the time-dependence of the scale factor of the
universe, which is governed by the semiclassical Einstein equation.
We can derive an expression for the influence functional in terms of the
Bogolubov coefficients as a function of the scale factor. The equation of
motion
becomes an Einstien-Langevin equation, from which
a new, extended theory of semiclassical gravity is obtained. This,
in our opinion, is necessary for furthering the investigation of quantum and
statistical processes in curved spacetimes. We are currently  pursuing
these investigations from this viewpoint.\\
c) {\it A fluctuation-dissipation relation} for non-equilibrium quantum fields.
Sciama \cite{Sciama} first suggested that the thermal radiance in a
uniformly accelerated observer (Unruh effect) and in black holes (Hawking
effect)
can be understood in terms of a fluctuation-dissipation relation.
This relation was also later derived for de Sitter spacetime via
linear response theory by Mottola \cite{Mottola}. These familiar cases all
deal with spacetimes with event horizons and thermal particle creation.
{}From earlier particle creation- backreaction studies in semiclassical gravity
\cite{cosbkr}
a general FDR was conjectured by one of us \cite{HuPhysica}
for quantum fields in  curved spacetimes.
It corresponds to a non-equilibrium generalization
of Hawking-Unruh effect to general dynamical spacetimes without event horizons.
Such a relation can in principle be identified from the results of this paper.
The interpretation of backreaction processes in terms of
fluctuation-dissipation relations
will be explored further in  \cite{HuSin94,HM3}.\\

\noindent {\bf Acknowledgement}
BLH thanks Sukanya Sinha for discussions at the preliminary stage of this work,
which was done and completed when AM visited the Maryland relativity theory
group in 92-93. Research is supported in part by the National Science
Foundation
under grant PHY91-19726.

\newpage
\appendix
\section{Influence Functional}
Here we describe the calculation of the influence functional.
 From (2.9) the influence functional is
$$
{\cal F}[x,x']=Tr[\hat{U}(t,t_i)\hat{\rho}_b (t_i)\hat{U}'^{\dag}(t,t_i)]
\eqno(A.1)
$$
where $\hat{U}(t)$ and $\hat{U}'(t)$ are the quantum propagators for the
actions \\ $S_E[q]+S_{int}[x(s),q]$ and $S_E[q]+S_{int}[x'(s),q]$, and
$x(s)$ and $x'(s)$ are treated as time dependent classical forcing terms.

Our first task is to determine the propagator for the action
$$
\eqalign{
S_E[q]+S_{int}[x(s),q]=\int\limits_{t_i}^tds \Biggl[
    \sum_n&\Bigl\{  {1\over 2}m_n(s)
   \Bigl(\dot q_n^2+b_n(s)q_n\dot{q}_n- \omega^2_n(s)q_n^2\Bigr) \Bigr\}\cr
& +    \sum_n\Bigl(-c_{1n}(s)F(x(s))q_n-c_{2n}(s)F(\dot{x}(s))q_n \cr
& -c_{3n}(s)F(x(s))\dot{q}_n(s)-c_{4n}(s)F(\dot{x}(s))\dot{q}_n\Bigr)\Biggr].
\cr}
\eqno(A.2)
$$
This interaction is the most general interaction possible which is linear in
the bath.
Dropping the $n$ subscript the Lagrangian for a mode takes the form
$$
\eqalign{
L(t)=\frac{1}{2}m(t)\Bigl(\dot{q}^2+b(t)q\dot{q}-\omega^2(t)q^2\Bigr)
&-q[c_1(t)F(x(t))+c_2(t)F(\dot{x}(t))] \cr
&-\dot{q}[c_3(t)F(x(t))+c_4(t)F(\dot{x}(t))]. \cr}
\eqno(A.3)
$$
Defining the canonical momenta the usual way we find that
$$
p_c=\frac{\partial L(t)}{\partial \dot{q}}=m(t)\dot{q}+m(t)b(t)\frac{q}{2}
-[c_3(t)F(x(t))+c_4(t)F(\dot{x}(t))].
\eqno(A.4)
$$
The Hamiltonian, $H(t)=p_c\dot{q}-L(t)$ then takes the form
$$
\eqalign{
H(t)= & \frac{p_c^2}{2m(t)}-\frac{b(t)}{4}(p_cq+qp_c)
+\frac{m(t)}{2}\left(\omega^2(t)+\frac{b^2(t)}{4}\right)q^2 \cr
&
+\Bigl[c_1(t)F(x(t))+c_2(t)F(\dot{x}(t))-\frac{b(t)}{2}(c_3(t)F(x(t))+c_4(t)F(\dot{x}(t)))\Bigr]q \cr
&+\frac{[c_3(t)F(x(t))+c_4(t)F(\dot{x}(t))]}{m(t)}p_c+
\frac{[c_3(t)F(x(t))+c_4(t)F(\dot{x}(t))]^2}{2m(t)}. \cr}
\eqno(A.5)
$$
The system is quantized by promoting $q,p_c$ to operators obeying
$[\hat{q},\hat{p}_c]=i\hbar$. Then writing
$$
\hat{q}=\sqrt{\frac{\hbar}{2\kappa}}(\hat{a}+\hat{a}^{\dag}),\;\;\;
\hat{p}_c=i\sqrt{\frac{\hbar\kappa}{2}}(\hat{a}^{\dag}-\hat{a})
\eqno(A.6)
$$
we find that (A.5) becomes
$$
\hat{H}(t)=f(t)\hat{A}+f^*(t)\hat{A}^{\dag}+h(t)\hat{B}+d(t)\hat{a}+d^*(t)
\hat{a}^{\dag}+g(t)
\eqno(A.7)
$$
where $\hat{A}$ and $\hat{B}$ are defined in (B.2) and
$$
f(t)=\frac{\hbar}{2}\left(\frac{m(t)\omega^2(t)}{\kappa}
+\frac{m(t)b^2(t)}{4\kappa}-\frac{\kappa}{m(t)}+ib(t)\right)
\eqno(A.8)
$$
$$
h(t)=\frac{\hbar}{2}\left(\frac{\kappa}{m(t)}+\frac{m(t)\omega^2(t)}{\kappa}
+\frac{m(t)b^2(t)}{4\kappa}\right)
\eqno(A.9)
$$
$$
d(t)=\sqrt{\frac{\hbar}{2\kappa}}\Biggl[c_1(t)F(x(t))+c_2(t)F(\dot{x}(t))
-\Bigl(\frac{b(t)}{2}-i\frac{\kappa}{m(t)}\Bigr)[c_3(t)F(x(t))+c_4(t)F(\dot{x}(t))]\Biggr]
\eqno(A.10)
$$
$$
g(t)=\frac{[c_3(t)F(x(t))+c_4(t)F(\dot{x}(t))]^2}{2m(t)}.
\eqno(A.11)
$$

In appendix B we have derived the evolutionary operator
generated by the Hamiltonian of (A.7). It has the form
$$
\hat{U}(t,t_i)=\hat{S}(r,\phi)\hat{R}(\theta)\hat{D}(p)\exp\left[-\frac{pp^*}{2}
-\frac{i}{\hbar}\int_{t_i}^tg(s)ds+\int_{t_i}^tds\int_{t_i}^sds'\dot{p}(s)
\dot{p}^*(s')\right].
\eqno(A.12)
$$

 From the first two equations of (B.13) we see that the squeeze and
rotation operators do not depend on $x$.
Thus $\hat{S}=\hat{S}',\hat{R}=\hat{R}'$.
Using this fact, the unitiary nature of the operators in the propagator,
the cyclic trace rule and the identity \cite{sqst}
$$
\hat{D}(p)\hat{D}(p')=\hat{D}(p+p')\exp\left[\frac{1}{2}(pp'^*-p^*p')\right]
\eqno(A.13)
$$
we find that (A.1) becomes
$$
\eqalign{
{\cal F}[x,x']=
& Tr[\hat{\rho}_b(t_i)\hat{D}(p-p')]\exp\left[\frac{1}{2}(pp'^*-p^*p'
-pp^*-p'p'^*)\right]
\cr
&\times\exp\Biggl[\int_{t_i}^{t}ds\int_{t_i}^{s}ds'\Bigl[\dot{p}(s)\dot{p}^*(s')
+\dot{p}'^*(s)\dot{p}'(s')\Bigr]
-\frac{i}{\hbar}\int_{t_i}^tds[g(s)-g'(s)]\Biggr].\cr}
\eqno(A.14)
$$
Making use of the integral identity
$$
\int_{b}^{a}g(t)dt\int_{b}^{a}h(t)dt=\int_{b}^{a}\int_{b}^{t}[g(t)h(t')+g(t')
h(t)]dt'dt
\eqno(A.15)
$$
its possible to write

$$
\eqalign{
{\cal F}[x,x']=&Tr[\hat{\rho}_b(t_i)\hat{D}(p-p')]
\exp\Bigl[-\frac{i}{\hbar}\int_{t_i}^tds[g(s)-g'(s)]\Bigr] \cr
& \times\exp\Biggl[\frac{1}{2}\int_{t_i}^{t}
ds\int_{t_i}^{s}ds'\Bigl([\dot{p}(s)-\dot{p}'(s)]
[\dot{p}'^*(s')+\dot{p}^*(s')] \cr
&+[\dot{p}(s')+\dot{p}'(s')][\dot{p}'^*(s)-\dot{p}^*(s)]\Bigr)\Biggr].\cr}
\eqno(A.16)
$$

We will now evaluate the influence functional for a squeezed thermal initial
state. Our first task is to compute the trace in (A.16).
Our initial state is of the form
$$
\hat{\rho}_b(t_i)=\hat{S}(r,\phi)\hat{\rho}_{th}
\hat{S}^{\dag}(r,\phi)
\eqno(A.17)
$$
where $\hat{\rho}_{th}$ is a thermal
density matrix of temperature $T$ defined by
the thermal density matrix takes the form
$$
\hat{\rho}_{th}=\left[1-\exp\left(\frac{-\hbar\omega}{k_BT}\right)\right]
\sum_n
\exp\left(\frac{-n\hbar\omega}{k_BT}\right)|n\rangle\langle n|
\eqno(A.18)
$$
and $\hat{S}(r,\phi)$
is a squeeze operator defined in (B.12).

The trace in (A.16) becomes
$$
Tr[\hat{\rho}_{b}(t_i)\hat{D}(p-p')]=
Tr[\hat{\rho}_{th}\hat{S}^{\dag}(r,\phi)\hat{D}(p-p')
\hat{S}(r,\phi)].
\eqno(A.19)
$$
Making use of \cite{sqst}
$$
\hat{S}^{\dag}(r,\phi)\hat{D}(p)\hat{S}(r,\phi)=\hat{D}(p\cosh r +
p^*\sinh re^{2i\phi})
\eqno(A.20)
$$
equation (B.16) and \cite{mes}
$$
Tr[\hat{\rho}_{th}\exp(t\hat{a}^{\dag}+u\hat{a})]=\exp\left[\frac{tu}{2}
\coth\left(\frac{\hbar\omega}{2k_BT}\right)\right]
\eqno(A.21)
$$
we find that
$$
Tr[\hat{\rho}_{b}(t_i)\hat{D}(p-p')]=
\exp\left[-\frac{1}{2}\coth\left(\frac{\hbar\omega}{2k_BT}
\right)|(p-p')\cosh r +
(p-p')^*\sinh r e^{2i\phi}|^2\right].
\eqno(A.22)
$$
Making use of the integral identity (A.15) we can write
$$
\eqalign{
Tr[\hat{\rho}_{b}(t_i)\hat{D}(p-p')]=&
\exp\Bigg[-\frac{1}{2}\coth\left(\frac{\hbar\omega}{2k_BT}\right)
\int_{t_i}^{t}ds\int_{t_i}^{s}ds' \cr
& \times\biggl\{[\dot{p}(s)-\dot{p}'(s)]
[\dot{p}(s')-\dot{p}'(s')]^*\cosh 2r \cr
&+[\dot{p}(s)-\dot{p}'(s)]^*
[\dot{p}(s')-\dot{p}'(s')]\cosh 2r \cr
& +\sinh 2r e^{-2i\phi}[\dot{p}(s)-\dot{p}'(s)]
[\dot{p}(s')-\dot{p}'(s')] \cr
& + \sinh 2r e^{2i\phi}[\dot{p}(s)-\dot{p}'(s)]^*
[\dot{p}(s')-\dot{p}'(s')]^*
\biggl\} \Bigg].\cr}
\eqno(A.23)
$$
Now put into (B.17)
$$
d=uF(x)+vF(\dot{x})
\eqno(A.24)
$$
where from (A.10)
$$
u=\sqrt{\frac{\hbar}{2\kappa}}\left(c_1-b\frac{c_3}{2}-i\kappa\frac{c_3}{m}
\right),\;\;
v=\sqrt{\frac{\hbar}{2\kappa}}\left(c_2-b\frac{c_4}{2}-i\kappa\frac{c_4}{m}
\right)
\eqno(A.25)
$$
and further define
$$
U=u\b^*+u^*\a^*,\;\;\;V=v\b^*+v^*\a^*.
\eqno(A.26)
$$
Using (B.17,A.23) and (A.16)
we find that the influence functional for a mode $n$ takes the form
$$
\eqalign{
F_n[x,x']=\exp \biggl[
& -{2i\over\hbar}\int\limits_{t_i}^tds\int\limits_{t_i}^{s}ds'
  \Bigl[\D(s)\mu_{1n}(s,s')\S(s') + \dot \D(s)\mu_{2n}(s,s')\S(s') \cr
& + \D(s)\mu_{3n}(s,s')\dot\S(s')+\dot\D(s)\mu_{4n}(s,s')\dot\S(s')\Bigr] \cr
& -{1\over\hbar}\int\limits_{t_i}^tds\int\limits_{t_i}^{s}ds'
  \Bigl[ \D(s)\nu_{1n}(s,s')\D(s')+ \D(s)\nu_{2n}(s,s')\dot\D(s') \cr
& +\dot\D(s)\nu_{3n}(s,s')\D(s') +\dot\D(s)\nu_{4n}(s,s')\dot\D(s') \Bigr] \cr
& -{i\over \hbar}\int_{t_i}^tds [g(s)-g'(s)] \biggr] \cr}
                           \eqno(A.27)
$$
where
$$
 \D (s)= [F(x(s))-F(x'(s))], ~~~  2\S(s') = [F(x(s'))+F(x'(s'))]
$$
$$
\dot{\D}(s)= [F(\dot{x}(s))-F(\dot{x}'(s))], ~~~  2\dot{\S}(s') =
[F(\dot{x}(s'))+F(\dot{x}'(s'))]
$$
$$
\eqalign{
& \mu_{1n}(s,s')=\frac{i}{2\hbar}[U(s)U^*(s')-U^*(s)U(s')] \cr
& \mu_{2n}(s,s')=\frac{i}{2\hbar}[V(s)U^*(s')-V^*(s)U(s')] \cr
& \mu_{3n}(s,s')=\frac{i}{2\hbar}[U(s)V^*(s')-U^*(s)V(s')] \cr
& \mu_{4n}(s,s')=\frac{i}{2\hbar}[V(s)V^*(s')-V^*(s)V(s')] \cr}
\eqno(A.28)
$$
and
$$
\eqalign{
 \nu_{1n}(s,s')=& \frac{1}{2\hbar}\coth\left(\frac{\hbar\omega}{2k_BT}\right)
\Bigl[\cosh 2r (U(s)U^*(s')+U^*(s)U(s')) \cr
&-\sinh 2re^{-2i\phi}U(s)U(s')-\sinh 2re^{2i\phi}U^*(s)U^*(s')\Bigr] \cr}
\eqno(A.29)
$$
$$
\eqalign{
\nu_{2n}(s,s')=& \frac{1}{2\hbar}\coth\left(\frac{\hbar\omega}{2k_BT}\right)
\Bigl[\cosh 2r (U(s)V^*(s')+U^*(s)V(s')) \cr
&-\sinh 2re^{-2i\phi}U(s)V(s')-\sinh 2re^{2i\phi}U^*(s)V^*(s')\Bigr] \cr}
\eqno(A.30)
$$
$$
\eqalign{
\nu_{3n}(s,s')=& \frac{1}{2\hbar}\coth\left(\frac{\hbar\omega}{2k_BT}\right)
\Bigl[\cosh 2r (V(s)U^*(s')+V^*(s)U(s')) \cr
&-\sinh 2re^{-2i\phi}V(s)U(s')-\sinh 2re^{2i\phi}V^*(s)U^*(s')\Bigr] \cr}
\eqno(A.31)
$$
$$
\eqalign{
\nu_{4n}(s,s')=& \frac{1}{2\hbar}\coth\left(\frac{\hbar\omega}{2k_BT}\right)
\Bigl[\cosh 2r (V(s)V^*(s')+V^*(s)V(s')) \cr
&-\sinh 2re^{-2i\phi}V(s)V(s')-\sinh 2re^{2i\phi}V^*(s)V^*(s')\Bigr] \cr}
\eqno(A.32)
$$
and
$$
g(s)=\frac{\Bigl[c_3(s)F(x(s))+c_4(s)F(\dot{x}(s))\Bigr]^2}{2m(s)}.
\eqno(A.33)
$$
Note that the $g$ term in the influence functional can be absorbed into the
system Lagrangian.
Of course the total influence functional is an infinite product of
influence functionals over n. Therefore if we define the spectral densitry as
$$
I(\omega,s,s')=\sum_n\delta(\omega-\omega_n)\frac{c_n(s)c_n(s')}{2\kappa_n}
\eqno(A.34)
$$
we obtain the results of (2.18-19) where we have put
$c_2,c_3,c_4=0$.

\section{Propagator}
Consider the Hamiltonian
$$
\hat{H}(t)=f(t)\hat{A}+f^*(t)\hat{A}^{\dag}+h(t)\hat{B}+d(t)\hat{a}+d^*(t)
\hat{a}^{\dag}+g(t)
\eqno(B.1)
$$
where
$$
\hat{A}=\frac{\hat{a}^2}{2},\;\;\;\;\;\hat{A}^{\dag}=
\frac{\hat{a}^{\dag 2}}{2},\;\;\;\;\;\hat{B}=\hat{a}^{\dag}\hat{a}+1/2.
\eqno(B.2)
$$
and $[\hat{a},\hat{a}^{\dag}]=1$.
We want to find the propagator for this general time dependent system.
We make the ansatz
$$
\hat{U}(t,t_i)=e^{x(t)\hat{B}}e^{y(t)\hat{A}}e^{z(t)\hat{A}
^{\dag}}e^{q(t)\hat{a}}e^{p(t)\hat{a}^{\dag}}e^{r(t)}.
\eqno(B.3)
$$
It has proved by Fernandez \cite{Fer} that this is global.
It must satisfy the evolution equation for the propagator
$$
\hat{H}(t)\hat{U}(t,t_i)=i\hbar\frac{\partial}{\partial t}\hat{U}(t,t_i)
\eqno(B.4)
$$
subject to the initial condition $\hat{U}(t_i,t_i)=1$.
We find that the operators $\hat{A}, \hat{A}^{\dag}, \hat{B}, \hat{a},
\hat{a}^{\dag}$
satisfy the following commutation relations
$$
[\hat{A},\hat{A}^{\dag}]=\hat{B}=\hat{B}^{\dag},\;\;\;\;\;[\hat{A},\hat{B}]=
\hat{A},\;\;\;\;\;[\hat{A}^{\dag},\hat{B}]=-2\hat{A}^{\dag}
$$
$$
[\hat{a},\hat{A}^{\dag}]=\hat{a}^{\dag},\;\;\;
[\hat{a}^{\dag},\hat{A}]=-\hat{a}
$$
$$
[\hat{a},\hat{B}]=\hat{a},\;\;\;
[\hat{a}^{\dag},\hat{B}]=-\hat{a}^{\dag} \eqno(B.5)
$$
$$
[\hat{a},\hat{A}]=[\hat{a}^{\dag},\hat{A}^{\dag}]=0.
$$
Making use of the commutation relations and the operator relation
$$
e^{u\hat{O}}\hat{P}e^{-u\hat{O}}=\hat{P}+u[\hat{O},\hat{P}]+\frac{u^2}{2!}
[\hat{O},[\hat{O},\hat{P}]]+...
\eqno(B.6)
$$
we find
$$
\eqalign{
e^{q\hat{a}}\hat{a}^{\dag}&=(\hat{a}^{\dag}+q)e^{q\hat{a}} \cr
e^{z\hat{A}^{\dag}}\hat{a}&=(\hat{a}-\hat{a}^{\dag}z)e^{z\hat{A}^{\dag}}
\cr
e^{y\hat{A}}\hat{a}^{\dag}&=(\hat{a}^{\dag}+y\hat{a})e^{y\hat{A}} \cr
e^{x\hat{B}}\hat{a}&=e^{-x}\hat{a}e^{x\hat{B}} \cr
e^{x\hat{B}}\hat{a}^{\dag}&=e^{x}\hat{a}^{\dag}e^{x\hat{B}}  \cr
e^{x\hat{B}}\hat{A}&=e^{-2x}\hat{A}e^{x\hat{B}} \cr
e^{y\hat{A}}\hat{A}^{\dag}&=(\hat{A}^{\dag}+\hat{B}y+y^2
\hat{A})e^{y\hat{A}} \cr
e^{x\hat{B}}\hat{A}^{\dag}&=
e^{2x}\hat{A}^{\dag}e^{x\hat{B}}. \cr}
\eqno(B.7)
$$
Substituting (B.3) into (B.4) and using (B.7) we find
$$
{\eqalign{
f&=i\hbar(\dot{y}e^{-2x}+\dot{z}y^2e^{-2x}) \cr
f^*&=i\hbar(\dot{z}e^{2x}) \cr
h&=i\hbar(\dot{x}+\dot{z}y) \cr
d&=i\hbar(\dot{q}(1-yz)e^{-x}+\dot{p}ye^{-x}) \cr
d^*&=i\hbar(\dot{p}e^x-\dot{q}ze^x) \cr
g&=i\hbar(\dot{p}q+\dot{r}). \cr}}
\eqno{B.8}
$$
Since the first three equations of (B.8) are independent of $d$ and $g$
the first three terms in the
propagator (B.3) are independent of the last three.
As it stands (B.3) is not necessarily unitary. Thus $x,y,z$ must satisfy some
further restrictions. If we write
$$
x=ln\alpha,\;\;\;\;\;y=-\beta\alpha,\;\;\;\;\;z=\beta^*/\alpha
\eqno(B.9)
$$
where
$$
\alpha=e^{-i\theta}\cosh r,\;\;\;\;\;\beta=-e^{-2i\varphi}\sinh r
\eqno(B.10)
$$
then we can write (B.3) as (using relations in \cite{sqst})
$$
\hat{U}(t,t_i)=\hat{S}(r,\phi)\hat{R}(\theta)e^{q\hat{a}}e^{p\hat{a}^{\dag}}e^r
\eqno(B.11)
$$
where $2\phi=2\varphi-\theta$ and
$$
\hat{R}(\theta)=e^{-i\theta \hat{B}},\;\;\;\;\;
\hat{S}(r,\phi)=\exp [r(\hat{A}e^{-2i\phi}-\hat{A}^{\dag}e^{2i\phi})].
\eqno(B.12)
$$
$\hat{S}$ and $\hat{R}$ are called squeeze and rotation operators
respectively \cite{sqst}. They are are both unitary as is required.
Substituting (B.9) into (B.8) we find
$$
\eqalign{
\hbar\dot{\alpha}&=-if^*\beta-ih\alpha \cr
\hbar\dot{\beta}&=ih\beta+if\alpha \cr
\hbar\dot{p}&=-i(d\beta^*+d^*\alpha^*) \cr
\dot{q}&=-\dot{p}^* \cr
\hbar\dot{r}&=-ig-\hbar\dot{p}q=-ig+\hbar\dot{p}p^*. \cr}
\eqno(B.13)
$$
The first 2 equations of (B.13) completely determine $\a$ and $\b$. The last
three determine $p,q,r$.
Making use of
$$
e^{\hat{F}+\hat{G}}=e^{\hat{F}}e^{\hat{G}}e^{-\frac{[\hat{F},\hat{G}]}{2}}
\eqno(B.14)
$$
where $\hat{F}$ and $\hat{G}$ are any operators that satisfy
$[\hat{F},\hat{G}]=$constant, we find that (B.11) becomes
$$
\hat{U}(t,t_i)=\hat{S}(r,\phi)\hat{R}(\theta)\hat{D}(p)e^{-pp^*/2}e^r
\eqno(B.15)
$$
where
$$
\hat{D}(p)=\exp[p\hat{a}^{\dag}-p^*\hat{a}]
\eqno(B.16)
$$
and
$$
p(t,t_i)=-\frac{i}{\hbar}\int_{t_i}^{t}dt[d(t)\beta^*(t)+d^*(t)\alpha^*(t)]
\eqno(B.17)
$$
$$
r(t,t_i)=-\frac{i}{\hbar}\int_{t_i}^{t}g(t)dt
+\int_{t_i}^{t}\dot{p}(t)p^*(t,t_i)dt.
\eqno(B.18)
$$
If we define
$$
\dot{p}(t)=\dot{p}_1(t)+i\dot{p}_2(t)
\eqno(B.19)
$$
and use the identity, (A.15),
we find that (B.15) becomes
$$
\eqalign{
\hat{U}(t,t_i)&=\hat{S}(r,\phi)\hat{R}(\theta)\hat{D}(p)
\exp\left[i\int_{t_i}^{t}
ds\int_{t_i}^{s}ds'[\dot{p}_2(s)\dot{p}_1(s')-\dot{p}_1(s)\dot{p}_2(s')]\right]
\cr
&\times\exp\left[\frac{-i}{\hbar}\int_{t_i}^{t}g(s)ds\right]. \cr}
\eqno(B.20)
$$
This form shows explicitly that the propagator is unitary.


\newpage
\begin{thebibliography}{999}

\bibitem {HPZ1}
B. L. Hu, J. P. Paz and Y. Zhang, Phys. Rev. {\bf D45}, 2843 (1992).

\bibitem {HPZ2}
B. L. Hu, J. P. Paz and Y. Zhang,
Phys. Rev. {\bf D47}, 1576  (1993).


\bibitem {FeyVer}
R. Feynman and F. Vernon, Ann. Phys. (NY) {\bf 24}, 118 (1963);
R. Feynman and A. Hibbs, {\it Quantum Mechanics and Path Integrals},
(McGraw - Hill, New York, 1965).

\bibitem  {CalLeg83}
A. O. Caldeira and A. J. Leggett, Physica {\bf 121A}, 587 (1983).

\bibitem  {Gra}
H. Grabert, P. Schramm and G. L. Ingold, Phys. Rep. {\bf 168}, 115 (1988).

\bibitem {HuWaseda}
B. L. Hu, ``Quantum Statistical Processes in the Early Universe''
in {\it Quantum Physics and the Universe}, Proc. Waseda Conference, Aug. 1992
ed. Namiki, K. Maeda, et al  (Pergamon Press, Tokyo, 1993).
Vistas in Astronomy {\bf 37}, 391 (1993).

\bibitem{BirDav}
N. Birrell and P. W. C. Davies {\it Quantum Fields in Curved Spaces}
(Cambridge University Press, Cambridge, 1982);
S. A. Fulling, {\it Aspects of Quantum Field Theory in Curved Spacetime}
(Cambridge University Press, Cambridge, 1987).

\bibitem {qos}
See, e.g., K. Lindenberg and B. J. West,
{\it The Nonequilibrium Statistical Mechanics
of Open and Closed Systems} (VCH Press, New York, 1990).
U. Weiss, {\it Quantum Dissipative Systems} (World Scientific, Singapore,
1993).

\bibitem {Zhang}
Yuhong Zhang, Ph. D. Thesis, University of Maryland (1990)

\bibitem {HuErice}
B. L. Hu, ``Quantum and Statistical Effects in Superspace Cosmology''
in {\it Quantum Mechanics in Curved Spacetime}, ed. J. Audretsch
and V. de Sabbata (Plenum, London 1990).

\bibitem {HuTsukuba} B. L. Hu, ``Statistical Mechanics and Quantum Cosmology'',
in {\it Proc. Second International Workshop on Thermal Fields and Their
Applications}, eds. H. Ezawa et al (North-Holland, Amsterdam, 1991).

\bibitem {Sinha}
S. Sinha, Ph. D. Thesis, University of Maryland (1991)

\bibitem {PazSin}
J. P. Paz and S. Sinha, Phys. Rev. {\bf D44}, 1038 (1991);
{\it ibid} {\bf D45}, 2823 (1992).

\bibitem {HPS}
B. L. Hu, J. P. Paz and S. Sinha, ``Minisuperspace as a Quantum  Open System''
in {\it Directions in General Relativity  Vol. 1: Misner Festschrift}
eds B. L. Hu, M. P. Ryan and C. V. Vishveswara
(Cambridge University Press, Cambridge, 1993).

\bibitem {decQC}
For earlier work on decoherence in quantum cosmology, see, e.g.,
C. Kiefer, Clas. Q. Grav. {\bf 4}, 1369 (1987);
J. J. Halliwell, Phys. Rev. {\bf D39}, 2912 (1989);
T. Padmanabhan, {\it ibid.} 2924 (1989).

\bibitem {HuPhysica}
B. L. Hu, Physica {\bf A158}, 399 (1989).

\bibitem {HuBanff}
B. L. Hu, in Proc. Third International Workshop on Thermal Field Theory and
Applications,  CNRS Summer Institute, Banff, Aug. 1993.
ed. R. Kobes and G. Kunstatter (World Scientific, Singapore, 1994).

\bibitem {HuZhaUncer}
B. L. Hu and Y. Zhang, in Proc. Third International
Workshop on Quantum Nonintegrability, Drexel University, Philadelphia,
May 1992, eds J. M. Yuan,  D. H. Feng and G. M. Zaslavsky
(Gordon and Breach, Langhorne, 1993);
B. L. Hu and Yuhong Zhang, Mod. Phys. Lett. A (1993).

\bibitem {AndHal}
A. Anderson and J. J. Halliwell, Phys. Rev. {\bf D48}, 2753 (1993).

\bibitem {envdec}
W. H. Zurek, Phys. Rev. {\bf D24}, 1516 (1981); {\bf D26}, 1862 (1982);
in {\it Frontiers of Nonequilibrium Statistical Physics}, ed. G. T. Moore
and M. O. Scully (Plenum, N. Y., 1986); Physics Today {\bf 44}, 36 (1991);
E. Joos and H. D. Zeh, Z. Phys. {\bf B59}, 223 (1985).
A. O. Caldeira and A. J. Leggett, Phys. Rev. {\bf A31}, 1059 (1985).
W. G. Unruh and W. H. Zurek, Phys. Rev. {\bf D40}, 1071 (1989);
B. L. Hu, J. P. Paz and Y. Zhang, Phys. Rev. {\bf D45}, 2843 (1992);
{\bf D47}, 1576 (1993);
W. H. Zurek, J. P. Paz and S. Habib, Phys. Rev. Lett. {\bf 47}, 1187 (1993);
J. P. Paz,  S. Habib and W. H. Zurek, Phys. Rev. {\bf D47}, 488 (1993);
J. P. Paz and W. H. Zurek, Phys. Rev. {\bf D48}, 2728 (1993).

\bibitem {conhis}
R. B. Griffiths, J. Stat. Phys. {\bf 36}, 219 (1984);
R. Omn\'es, J. Stat Phys. {\bf 53}, 893, 933, 957 (1988);
Ann. Phys. (NY) {\bf 201}, 354 (1990); Rev. Mod. Phys. {\bf 64}, 339 (1992).

\bibitem{GelHar1}
M. Gell-Mann and J. B. Hartle, in
{\it Complexity, Entropy and the Physics of Information}, ed.
by W. H. Zurek (Addison-Wesley, Reading, 1990);
H. F. Dowker and J. J. Halliwell, Phys. Rev. {\bf D46}, 1580 (1992);
A. Albrecht, Phys. Rev. {\bf D48}, 3768 (1993);
J. Twamley, Phys. Rev. {\bf D48} (1993).

\bibitem{GelHar2}
M. Gell-Mann and J. Hartle, Phys. Rev. {\bf D47}, 3345 (1993);
T. Brun, Phys. Rev. {\bf D47}, 3383 (1993).

\bibitem{CalHuDCH}
E. Calzetta and B. L. Hu, ``Decoherence of Correlation Histories''
in {\it Directions in General Relativity, Vol II: Brill Festschrift}
eds B. L. Hu and T. A. Jacobson (Cambridge University Press, Cambridge, 1993).

\bibitem {ZurekPTP}
W. H. Zurek, Prog. Theor. Phys. {\bf 89}, 281 (1993).

\bibitem {HarLH}
J. B. Hartle, Les Houches 1992 Lectures (1993).
J. B. Hartle, ``Quantum Mechanics of Closed Systems"
in {\it Directions in General Relativity, Vol. 1: Misner Festschrift}
eds B. L. Hu, M. P. Ryan and C. V. Vishveswara
(Cambridge Univ., Cambridge, 1993).


\bibitem {Omnes}
R. Omn\'es, Rev. Mod. Phys. {\bf 64}, 339 (1992).

\bibitem {qsf1} B. L. Hu, J. P. Paz and Y. Zhang, ``Stochastic Dynamics of
Interacting Quantum Fields'' in preparation (1994).

\bibitem {HuBelgium}
B. L. Hu, J. P. Paz and Y. Zhang (1993) ``Quantum Origin of Noise and
Fluctuations in Cosmology'',
in {\it The Origin of Structure in the Universe}, edited by E. Gunzig and
P. Nardone (Kluwer, Dordrecht), p. 227.

\bibitem {HMLA}
B. L. Hu and A. Matacz, ``Quantum Noise in Gravitation and Cosmology''
in Proc. International Workshop on {\it Fluctuations and Order: A New
Synthesis},
Los Alamos, Sept. 1993, ed. Marko Millonas (MIT Press, Cambridge, 1994).

\bibitem {Haw75}
S. W. Hawking, Comm. Math. Phys. {\bf 43}, 199 (1975).

\bibitem {GibHaw}
G. Gibbons and S. W. Hawking, Phys. Rev. {\bf D15}, 2738 (1977).

\bibitem {Unr}
W. H. Unruh, Phys. Rev. {\bf D14}, 870 (1976).


\bibitem {KuoFor}
C. I. Kuo and L. H. Ford, Phys. Rev. {\bf D47}, 4510 (1993).

\bibitem {CalHuSG}
E. Calzetta and B. L. Hu,``Noise and Fluctuations in Semiclassical Gravity",
Univ. Maryland preprint 93-216 (1993).

\bibitem {HM3}
B. L. Hu and A. Matacz, ``Backreaction in
Semiclassical Cosmology: the Einstein-Langevin Equation",
Univ. Maryland preprint 94-31 (1993).

\bibitem {SinHu}
S. Sinha and B. L. Hu, Phys. Rev. {\bf D44}, 1028 (1991).

\bibitem {cgea}
B. L. Hu and Y. Zhang, ``Coarse-Graining, Scaling, and Inflation"
Univ. Maryland Preprint 90-186 (1990);
B. L. Hu, in {\it Relativity and Gravitation: Classical
and Quantum} Proc. SILARG VII, Cocoyoc, Mexico 1990.
eds. J. C. D' Olivo et al (World Scientific, Singapore 1991).

\bibitem{ctp}
J. Schwinger, J. Math. Phys. {\bf 2} (1961) 407;
L. V. Keldysh, Zh. Eksp. Teor. Fiz. {\bf 47 }, 1515 (1964)
[Engl. trans. Sov. Phys. JEPT {\bf 20}, 1018 (1965)].
G. Zhou, Z. Su, B. Hao and L. Yu, Phys. Rep. {\bf 118}, 1 (1985);
Z. Su, L. Y. Chen, X. Yu and K. Chou, Phys. Rev. {\bf B37}, 9810 (1988).
B. S. DeWitt, in {\it Quantum Concepts in Space and Time}
ed. R. Penrose and C. J. Isham (Claredon Press, Oxford, 1986);
R. D. Jordan, Phys. Rev. {\bf D33}, 44 (1986)
E. Calzetta and B. L. Hu, Phys. Rev. {\bf D35}, 495 (1987).


\bibitem {disQG}
E. Calzetta, Class. Quan. Grav. {\bf 6}, L227 (1989);
Phys. Rev. {\bf D 43}, 2498 (1991).
B. L. Hu, ``Quantum and Statistical Effects in Superspace
Cosmology'' in {\it Quantum Mechanics in Curved Spacetime}, ed. J. Audretsch
and V. de Sabbata (Plenum, London 1990).




\bibitem {inflation}  A. H. Guth, Phys. Rev. {\bf D23}, 347 (1981);
K. Sato, MNRAS {\bf 195}, 467 (1981);  A. D. Linde, Phys. Lett. {\bf 108B}, 389
(1982);
 A. Albrecht and P. J. Steinhardt, Phys. Rev. Lett. {\bf 48}, 1220 (1982).

\bibitem{stoinf}
A. A. Starobinsky, in {\it Field Theory, Quantum Gravity and
Strings}, ed. H. J. de Vega and N. Sanchez (Springer, Berlin 1986);
J. M. Bardeen and G. J. Bublik, Class. Quan. Grav. {\bf 4}, 473 (1987).

\bibitem {galforinf}
A. Guth and S. Y. Pi, Phys. Rev. Lett. {\bf 49}, 1110 (1982).
A. A. Starobinsky, Phys. Lett. {\bf 117B}, 175 (1982).
S. W. Hawking, Phys. Lett. {\bf 115B}, 295 (1982).
J. M. Bardeen, P. J. Steinhardt and M. S. Turner, Phys. Rev. {\bf D28}, 629
(1983);
R. Brandenberger, R. Kahn and W. Press, Phys. Rev. {\bf D 28}, 1809 (1983).
V. Mukhanov, H. Feldman and R. Brandenberger, Phys. Rep. {\bf 215}, 203 (1992).


\bibitem {GutPi}
A. H. Guth and S. Y. Pi, Phys. Rev. {\bf D32}, 1899 (1985).

\bibitem {CorBru}
J. M. Cornwall and R. Bruinsma, Phys. Rev. {\bf D38}, 3146 (1988)

\bibitem {decinf}
M. Sakagami, Prog. Theor. Phys {\bf 79}, 443 (1988);
R. Brandenberger, R. Laflamme and M. Mijic, Mod. Phys. Lett {\bf A5}, 2311
(1990); H.A. Feldman and A.Y. Kamenshchik, Class. Quant. Grav {\bf 8}, L65
(1991);
J.P. Paz, in {\em Proc. Second International Workshop on Thermal Fields
and Their Applications}, ed. H. Ezawa et al (North-Holland, Amsterdam, 1991);

\bibitem {Mat}
A. Matacz, Class. Quant. Grav {\bf 10}, 509 (1993);
R. Laflamme and A. Matacz, Int. J. Mod. Phys. D. {\bf 2}, 171 (1993);
A. Matacz, Phys. Rev. D. (1994).


\bibitem{cpc}
L. Parker, Phys. Rev. {\bf 183}, 1057 (1969); {\bf D3}, 346 (1971);
R. U. Sexl and H. K. Urbantke, { Phys. Rev.}, {\bf 179}, 1247 (1969)
Ya. B. Zel'dovich, Pis'ma Zh. Eksp. Teor. Fiz, {\bf 12} ,443 (1970)
[JETP Lett. {\bf 12}, 307(1970)];
Ya. B. Zel'dovich and A. A. Starobinsky,
Zh. Teor. Eksp. Fiz. {\bf 61} 2161, (1971) [Sov. Phys. JETP
{\bf 34}, 1159 (1972)].

\bibitem{cosbkr}
B. L. Hu  and L. Parker, Phys. Rev. {\bf D17}, 933 (1978).
J. Hartle and B. L. Hu, Phys. Rev. {\bf D20}, 1772 (1979).
E. Calzetta and B. L. Hu,
Phys. Rev. {\bf D35}, 495 (1987).

\bibitem {dissip}
E. Calzetta and B. L. Hu,
Phys. Rev. {\bf D40}, 656 (1989);
J. P. Paz, Phys. Rev. {\bf D41}, 1054 (1990).

\bibitem  {Ang}
J. R. Anglin, Phys. Rev. {\bf D47}, 4525 (1993).

\bibitem {milburn}
M. A. Dupertuis and S. Stenholm, J. Opt. Soc. Am. {\bf B4}, 1102 and 1124
(1987);
A. S. Parkins and C. W. Gardiner, Phys. Rev. {\bf A40}, 3796 (1989).
We thank Gerald Milburn for suggesting this reference.

\bibitem{gilson}
C.R. Gilson, S.M. Barnett and S.Stenholm, J. Mod. Opt {\bf 34}, 949 (1987)
and references therein.

\bibitem{sqth}
M. S. Kim, F.A.M. de Oliveira and P.L. Knight, Phys. Rev. {\bf A40}, 2494
(1989);
M. S. Kim and V. Buzek, Phys. Rev. {\bf A47}, 610 (1993) and references
therein.

\bibitem {sqst}
C. M. Caves and B. L. Schumacher, Phys. Rev. {\bf A31}, 3068, 3093 (1985);
B. L. Schumacher, Phys. Rep. {\bf 135}, 317 (1986).


\bibitem {GriSid}
L. Grishchuk and Y. V. Sidorov, Phys. Rev. {\bf D42}, 3414 (1990).

\bibitem{HKM}
B. L. Hu, G. W. Kang and A. Matacz, Int. J. Mod. Phys. A (1993).

\bibitem {PazSpain}
J. P. Paz, in ``The Physical Origin of Time Asymmetry'' ed. J. J. Halliwell et
al
(Cambridge Univ. Press, 1994).

\bibitem {HabKan}
S. Habib and H. E. Kandrup, Phys. Rev. {\bf D46}, 5303 (1992);
S. Habib, in {\it Stochastic Processes in Astrophysics} (1993).

\bibitem {CalMaz}
E. Calzetta and D. Mazzitelli, Phys. Rev. {\bf D42} 4066 (1990).

\bibitem  {fdr}
H. Callen and T. Welton, Phys. Rev. {\bf 83}, 34 (1951);
M. S. Green, J. Chem. Phys. {\bf 19}, 1036 (1951);
R. Kubo, J. Phys. Soc. Japan {\bf 12}, 570 (1957); Rep. Prog. Phys. {\bf 29},
255 (1966).

\bibitem {Sciama}
D. W. Sciama, ``Thermal and Quantum Fluctuations in Special
and General Relativity: An Einstein Synthesis''
in {\it Centenario di Einstein} (Editrici Giunti Barbera Universitaria, 1979);
P. Candelas and D. W. Sciama, Phys. Rev. Lett. 38, 1372 (1977)

\bibitem {Mottola}
E. Mottola,
Phys. Rev. {\bf D33}, 2136 (1986).

\bibitem  {HuSin94}
B. L. Hu and S. Sinha, ``Fluctuation-Dissipation Relation in Cosmology",
   Univ. Maryland preprint pp93-164 (1993).

\bibitem  {HakAmb}
V. Hakim and V. Ambegoakar, Phys. Rev. {\bf A32}, 423 (1985).

\bibitem {LegRMP}
A. J. Leggett, S. Chakravrty, A. T. Dorsey,  M. P. A. Fisher, A. Garg
and W. Zwerger, Rev. Mod. Phys. {\bf 59},  No. 1 (1987) 1.

\bibitem {Dekker}
H. Dekker, Phys. Rev. {\bf A16}, 2126 (1977).

\bibitem {Sim}
J. Z. Simon, Private Communication

\bibitem {Fer}
F.M. Fernandaz, Phys. Rev. {\bf A40}, 41 (1989).

\bibitem {mes}
A. Messiah, {\it Quantum Mechanics Vol 1}, page 451 (North-Holland, Amsterdam,
1961).


\end{thebibliography}
\end{document}